\pdfoutput=1
\documentclass[useAMS,usenatbib]{mn2e}
\usepackage{epsfig,amsmath,amssymb,amsfonts,mathrsfs,latexsym,graphicx}
\begin{document}

\title[Characteristic Velocities of Stripped-Envelope Core-Collapse Supernova
  Cores]{Characteristic Velocities of Stripped-Envelope Core-Collapse Supernova Cores$^{*}$}

\author[Maurer et al.]{J. I. Maurer$^{1,**}$, P. A. Mazzali$^{1,2,3}$, J. Deng$^4$,
  A. V. Filippenko$^5$, M. Hamuy$^6$, \and R. P. Kirshner$^7$, T. Matheson$^8$, 
M. Modjaz$^5$, E. Pian$^{9,10}$, M. Stritzinger$^{11,12}$, 
\and S. Taubenberger$^{1}$, S. Valenti$^{10}$
\\$^1$ Max Planck Institut f\"ur
  Astrophysik, Karl-Schwarzschild-Str.1, 85741 Garching, Germany
\\$^2$ Scuola Normale Superiore, Piazza dei Cavalieri, 7, 56126 Pisa, Italy
\\$^3$ National Institute for Astrophysics-OAPd, Vicolo
dell'Osservatorio, 5, 35122 Padova, Italy
\\$^4$ National Astronomical Observatories, CAS, 20A Datun Road,
Chaoyang District, Beijing 1000012, China
\\$^5$ Department of Astronomy, University of California, Berkeley, CA 94720-3411, USA
\\$^6$ Departamento de Astronomia, Universidad de Chile, Casilla 36-D,
Santiago, Chile
\\$^7$ Harvard-Smithsonian Center for Astrophysics, 60 Garden Street,
Cambridge, MA 02138, USA
\\$^8$ National Optical Astronomy Observatory, 950 N. Cherry Avenue,
Tucson, AZ 85719-4933, USA
\\$^9$ INAF, Trieste Astronomical Observatory, Via G. B. Tiepolo, 11,
1-34143 Trieste, Italy
\\$^{10}$ European Southern Observatory, Karl-Schwarzschild-Strasse 2
D-85748 Garching bei M\"unchen, Germany
\\$^{11}$ Las Campanas Observatory, Carnegie Observatories, Casilla 601, La
Serena, Chile
\\$^{12}$ Dark Cosmology Centre, Niels Bohr Institute, University 
of Copenhagen, Juliane Maries Vej 30, 2100 Copenhagen \O, Denmark
\\
\\$^{*}$ Based on observations at ESO-Paranal, Prog. 081.D-0173(A), 082.D-0292(A)
\\$^{**} maurer@mpa$-$garching.mpg.de$
}

\maketitle

\begin{abstract}

The velocity of the inner ejecta of stripped-envelope core-collapse supernovae 
(CC-SNe) is studied by means of an analysis of their nebular spectra.
Stripped-envelope CC-SNe are the result of the explosion of bare 
cores of massive stars ($\geq 8$ M$_{\odot}$), and their late-time spectra are 
typically dominated by a strong [O {\sc i}] $\lambda\lambda$6300, 6363 emission line 
produced by the innermost, slow-moving ejecta which are not visible at earlier
times as they are located below the photosphere. 
A characteristic velocity of the inner ejecta is obtained for a sample of 56 
stripped-envelope CC-SNe of different spectral types (IIb, Ib, Ic) using direct 
measurements of the line width as well as spectral fitting. For most SNe, this 
value shows a small scatter around 4500 km s$^{-1}$.
Observations ($< 100$ days) of stripped-envelope CC-SNe have revealed
a subclass of very energetic SNe, termed broad-lined SNe (BL-SNe) or
hypernovae, which are characterised by broad absorption lines in the early-time
spectra, indicative of outer ejecta moving at very high velocity ($v \geq 0.1
c$).
SNe identified as BL in the early phase show large variations of core velocities
at late phases, with some having much higher and some having similar velocities
with respect to regular CC-SNe. This might indicate asphericity of the inner
ejecta of BL-SNe, a possibility we investigate using synthetic three-dimensional 
nebular spectra.

\end{abstract}


\maketitle

\section{Introduction}
\label{intro}

Massive stars ($>8$ M$_\odot$) collapse when the nuclear fuel in their
central regions is consumed, producing a core-collapse supernova (CC-SN) and
forming a black hole or a neutron star. CC-SNe with a H-rich spectrum are classified 
as Type II. If the envelope was stripped to some degree prior to the explosion, the
SNe are classified as Type IIb (strong He lines, and weak but clear H), Type Ib (strong
He lines but no H), and Type Ic (no He or H lines) \citep{1997ARA&A..35..309F}.

Some CC-SNe, called broad-lined SNe (BL-SNe), exhibit very broad absorption lines in
their early phase, resulting from the presence of sufficiently massive ejecta
expanding at high velocities. BL-SNe seem to be preferentially of Type Ic [two
exceptions are the Type IIb SN 2003bg \citep{2009arXiv0908.1783H} and the Type
Ib SN 2008D \citep{2008Sci...321.1185M,2009ApJ...702..226M}]; see also SN 1987K \citep{1988AJ.....96.1941F},
which is mentioned by \citep{2009arXiv0908.1783H}. Some BL-SNe can reach
kinetic energies
of $\ge$ 10$^{52}$ ergs. They are sometimes called hypernovae, and can be
associated with long-duration gamma-ray bursts (GRBs) \citep[see][and references
therein]{2006ARA&A..44..507W}. However not all BL-SNe are associated
with GRBs. 

An important question in the context of CC-SNe is
how the gravitational energy is converted into outward motion of the ejecta
during the collapse; see \citep{2007PhR...442...38J} for a recent review. In
GRB scenarios a relativistic outflow is launched by the central engine and
deposits some fraction of its energy into the SN ejecta, probably preferentially
along the polar axis, which might cause strong asymmetries
\citep[e.g.,][]{2002ApJ...565..405M}. The nearest, best-studied GRB-SNe are SN
1998bw / GRB 980425 \citep{1999A&AS..138..465G}, SN 2003dh / GRB 030329 
\citep{2004cetd.conf..351M}, SN
2003lw / GRB 031203 \citep{2004ApJ...609L...5M}, and SN 2006aj / GRB/XRF 060218
\citep{2006Natur.442.1011P}, although it is not fully established that the GRBs
(or X-ray flashes) accompanying nearby CC-SNe can be compared directly to 
high-redshift GRBs. CC-SNe may be characterised by asphericities although a jet 
does not necessarily form
\citep{2003ApJ...584..971B,2007ApJ...655..416B,2004ApJ...608..391K,2006MNRAS.370..501M,2009ApJ...691.1360T}.

Extremely massive stars ($> 100 $ M$_\odot$) are thought to end their lives as
pair-instability SNe (PI SNe). A star with enough mass to form a He core with
more than 40 M$_\odot$ will suffer electron-positron pair instability, leading
to rapid collapse. This triggers explosive oxygen burning, leading to the
complete disruption of the star \citep{1967PhRvL..18..379B,2005IAUS..228..297H}.
This process can produce large amounts of $^{56}$Ni. The ejecta mass and the
explosion kinetic energy are high, but the ejecta velocities are moderate
\citep{2005ApJ...633.1031S}.

Stripped CC-SNe, which lost part of their envelope before collapse, offer a clearer 
view of their inner ejecta than SNe which have retained it. Thus, in this paper we
exclusively address stripped CC-SNe (Types IIb, Ib, Ic); we do not include
Type II SNe, despite using the term ``CC-SN.''

Asphericities in their inner and outer ejecta \citep[e.g.,][]{2001ApJ...559.1047M}
are evident in at least some CC-SNe. Two main indicators are
velocity differences of Fe and lighter-element lines, and polarisation
measurements \citep[e.g.,][]{1991A&A...246..481H}.

Independent of their type, with time SNe become increasingly transparent to
optical light, as the ejecta thin out. At late times ($> 200$ days after the
explosion), the innermost layers of the SN can be observed. This epoch is called
the nebular phase, because the spectrum turns from being absorption dominated to
emission, mostly in forbidden lines.  In this phase the radiated energy of a SN
is provided by the decay of radioactive $^{56}$Co (which is produced by the earlier
decay of $^{56}$Ni). Decaying $^{56}$Co emits $\gamma$-rays and positrons which
are absorbed by the SN ejecta. As the deposition rate of $\gamma$-rays and
positrons depends on the density and $^{56}$Ni distribution, the inner parts of the
SN dominate the nebular spectra.

Recently, several authors
\citep{2008Sci...319.1220M,2008ApJ...687L...9M,2009arXiv0904.4632T} have studied
nebular spectra of CC-SNe. They concentrated on the shape of the [O {\sc i}]
$\lambda\lambda$6300, 6364 doublet (which is produced by much of the
mass) and concluded that torus-shaped oxygen distributions might cause the
double-peaked [O {\sc i}] profile observed in many CC-SNe nebular spectra. However,
there is ongoing discussion regarding whether geometry is the dominant reason 
for this type of line profile \citep[][also see Appendix A]{2009arXiv0904.4256M}.

Several authors have modelled nebular-phase spectra of SNe to derive quantities 
such as the $^{56}$Ni mass
\citep[e.g.,][]{2004ApJ...614..858M,2006A&A...460..793S,2006MNRAS.369.1939S,2007ApJ...658L...5M},
ejecta velocities \citep[e.g.,][]{2007Sci...315..825M}, asphericities
\citep[e.g.,][]{2005Sci...308.1284M,2007ApJ...670..592M,2008Sci...319.1220M}, and
elemental abundances \citep[e.g.,][]{2007ApJ...658L...5M, 2007ApJ...666.1069M}.
The nebular phase is especially suitable for studying the core of SNe. If
different explosion scenarios are involved for different types of CC-SNe, one
might expect the largest, most revealing differences to be in the central region of
the explosion. Therefore, in contrast to the standard classification of BL-SNe,
which is based primarily on early-time spectroscopy and describes the velocity field of
the outer SN layers, here we focus on the centre of the explosion. We have
modelled the nebular spectra of over 50 SNe, the largest sample of CC-SNe so
far, and obtained a statistically significant representation of their core
velocities.

We describe our data set in Section 2 and the modelling procedure in Section
3. In Section 4 we test the reliability of the modelling
approach. Results are discussed in Section 5.

\section{Data Set}
\label{ism1} 

We collected nebular spectra of 56 CC-SNe. This sample includes all the spectra
presented by \citet{2001AJ....121.1648M}, \citet{2008ApJ...687L...9M}, and
\citet{2009arXiv0904.4632T} for which a spectral fit was possible. The most
important criteria for selection were a reasonable signal-to-noise ratio and a
spectral coverage of at least the region between 6000 and 6500~\AA. Most spectra
range from 4000 to 10000~\AA, allowing modelling of the Fe-group, oxygen,
calcium, and carbon lines. If we found evidence for an underlying continuum, we
tried to remove it using a linear fit. When several nebular spectra were
available for a given SN, we chose the one closest to 200 days, although the precise
epoch has little influence as long as the spectrum is nebular, as shown in
Section 4.

Unfortunately, most of our spectra are not properly flux-calibrated. If an
estimate of the $^{56}$Ni mass was available in the literature for SNe with
uncalibrated spectra, we used these values (Table 1). For the remaining SNe 
with uncalibrated spectra we tried to estimate
the $^{56}$Ni mass from the light curve. We emphasise that the exact $^{56}$Ni
mass is not important for our study, as we show in Section 4.

\begin{table*}
\begin{tabular}{|l|c|c|c|c|c|c|}
\hline
SN  & log($L_{\rm Bol}$) & $M_{\rm Ni}$ (M$_\odot$) & $\beta$ & $M_{\rm ej,tot}$ ($M_\odot$)& $E_{\rm kin,tot}$ (10$^{51}$ ergs) & References ($L_{\rm peak}$, $M_{\rm Ni}$, $M_{\rm tot} \& E_{\rm tot}$)\\
\hline
1990I  & 42.4    & 0.11 $\pm$ 0.02    & 0.044 & 3.7 $\pm$ 0.7 & 1.1 $\pm$ 0.1 & 1,1,1\\
1993J  & 42.2    & 0.08 $\pm$ 0.02    & 0.050 & $\sim$3 & $\sim$1         & 2,2,13\\
1994I  & 42.2    & 0.065 $\pm$ 0.03   & 0.041 & $\sim$0.9 & $\sim$ 1.0& 3,3,7 \\
1997dq & 42.2    & 0.15 $\pm$ 0.03    & 0.095 & $\sim$8--10 & 10--20       & 5,5,5 \\
1997ef & 42.2    & 0.135 $\pm$ 0.025  & 0.085 & $\sim$8--10 & 10--20       & 4,5,5\\
1998bw & 42.8    & 0.49 $\pm$ 0.04    & 0.079 & $\sim$14  &  $\sim$60  & 3,3,15 \\
2002ap & 42.1    & 0.073 $\pm$ 0.02   & 0.058 & 2.5--5 & 4--10 & 3,3,14 \\
2003jd & 42.8    & 0.36 $\pm$ 0.04    & 0.057 & 3 $\pm$ 0.5 & 7$^{+3}_{-2}$& 3,3,3\\
2004aw & 42.4    & 0.21 $\pm$ 0.03    & 0.084 & 3.5--8.0 & 3.5--9.0 & 3,16,16 \\
2006aj & 42.7    & 0.20 $\pm$ 0.04    & 0.040 & $\sim$2 & $\sim$2 & 3,12,12  \\
2007Y  & 42.1    & 0.06 $\pm$ 0.01    & 0.048 & $\sim$0.5 & $\sim$0.1 & 8,8,8 \\
2007gr & 42.2    & 0.08 $\pm$ 0.02    & 0.046 & 1.5--3 & 1.5--3 & 10,11\\
2007ru & 42.9    & 0.4                & 0.045 & $1.3^{+1.1}_{-0.8}$ & $5^{+4.7}_{-3.0}$ & 9,9,9 \\
2008D  & 42.2    & 0.09 $\pm$ 0.02    & 0.057 & $\sim$7& $\sim$6 & 4,4,4\\
2008ax & 42.3    & 0.08 $\pm$ 0.02    & 0.040 & 3--6 & $\sim$1 & 6,6,6 \\
\hline
\label{tab1}
\end{tabular}
\caption{Peak luminosities, $^{56}$Ni masses, the ratio $\beta$ $\equiv$
  $M_{\rm Ni}$ (M$_\odot$)/$L_{\rm Bol,42}$, the total ejecta
  mass, and the total kinetic energy for 15 CC-SNe taken from the
  literature (see references). Typically $\beta \approx 0.06$ with a
  dispersion of roughly 70\%. References: $^1$\citet{2004A&A...426..963E}, 
  $^2$\citet{1994A&A...281L..53B}, $^3$\citet{2008MNRAS.383.1485V},
  $^4$\citet{2008Sci...321.1185M}, $^5$\citet{2004ApJ...614..858M},
  $^6$\citet{2008MNRAS.389..955P}, $^7$\citet{2006MNRAS.369.1939S}, 
  $^{8}$\citet{2009ApJ...696..713S}, $^{9}$\citet{2009ApJ...697..676S},
  $^{10}$ Hunter, in prep., $^{11}$\citet{2008ApJ...673L.155V}, 
  $^{12}$\citet{2006Natur.442.1018M}, $^{13}$\citet{1993Natur.364..507N},
  $^{14}$\citet{2002ApJ...572L..61M}, $^{15}$\citet{2000IAUS..195..347N},
  $^{16}$\citet{2006MNRAS.371.1459T}.}
\end{table*}

For several SNe only one light-curve point exists (generally the detection
magnitude in the $V$ or $B$ bands, or unfiltered). To obtain a (very crude)
estimate of the $^{56}$Ni mass from this single data point we employed the
following procedure. For the SNe with known $^{56}$Ni mass and 
bolometric peak luminosity, the ratio
of bolometric peak luminosity (in units of 10$^{42}$ ergs) to $^{56}$Ni mass (in
units of M$_\odot$) can be calculated. This ratio varies between 0.040 and
0.095, with a rather uniform distribution (see Table 1), which reflects the
variety of light-curve widths of different SNe (e.g., SNe with broader light
curves have more $^{56}$Ni for the same peak luminosity). From the estimates of
the $^{56}$Ni mass and the peak luminosity of these SNe, we can derive a
relation

\begin{equation}
M_{Ni}({\rm M}_\odot) \approx 0.058^{+0.042}_{-0.025}  \, L_{\rm peak,42}.
\label{form1}
\end{equation}

This estimate should be compared with a similar one obtained for SNe~Ia by
\citet{2006A&A...460..793S}, who found $M_{\rm Ni}$ (M$_\odot$) $= 0.050 \, 
L_{\rm peak,42}$. The difference probably arises from the different
densities and compositions of SNe~Ia and CC-SNe.

To estimate the peak luminosity from the measured magnitude, we first tried to
determine the bolometric luminosity at the time of detection. For some SNe an
estimate for both the Milky Way and host-galaxy absorption is available in the
literature. For most SNe, however, only the former is known 
\citep{1998ApJ...500..525S}. In this case we assume a host-galaxy absorption
between 0 and 1.0 mag (unfiltered) and treat this range as an uncertainty
affecting our estimate. If the Milky Way absorption is not known as
well, we assume an uncertainty between 0 and 1.5 mag (unfiltered). Unfiltered magnitudes are treated as bolometric,
$V$-band magnitudes are converted to bolometric magnitudes using $m_{\rm Bol}
= m_V - 0.3 \pm 0.2$ , and $B$-band magnitudes are converted using $m_{\rm Bol} =
m_B - 0.8 \pm 0.4$. For the distance moduli and the errors in the distance we
took the values listed in
NED\footnote[1]{$http://nedwww.ipac.caltech.edu/forms/byname.html$} for the SN
host galaxies (Virgo+GA+Shapley). We then estimated the epoch of detection
(which is close to maximum light for most of the SNe of our sample) and the
uncertainty in this value from the references given in Table 2. Comparing to
light curves of well-observed SNe we determined the value and uncertainty of the
peak luminosity, including uncertainties related to absorption, conversion from
filtered to bolometric luminosity, and the lack of a well-sampled light curve.
Combining this estimate with Equation (\ref{form1}), we obtained $^{56}$Ni masses
for all 56 SNe of our sample. These are listed in Tables 1 and 2. The possible
error of this method is very large, spanning roughly a factor of 20. This uncertainty
estimate is very conservative; for most SNe the actual error should be much
smaller. However, it is sufficient for our purposes (as we show in the Section
4).

\begin{table*}
\begin{tabular}{|l|c|c|c|c|c|c|c|}
\hline
SN & $m$ & $A_V$ (mag) & $m - M$ (mag) & $\Delta d$ & log($L_{\rm peak}$) & $M_{\rm Ni}$ & References ($m$, $A_V$) \\
\hline
1983N  & 11.3$_V$ & 0.51 $\pm$ 0.05 & 28.02 $\pm$ 0.3 & 0 $\pm 5$ & 42.45$^{+0.31}_{-0.22}$ & 0.17$^{+0.43}_{-0.11}$ & 31,31\\
1985F  & 12.1$_B$ & 0.70 & 29.72 & 0 $\pm 4$  & 43.08$^{+0.33}_{-0.25}$ & 0.73$^{+1.91}_{-0.49}$ & 1,2\\
1987M  & -          & -    & -     & -          &43.00$^{+0.18}_{-0.18}$ & 0.60$^{+0.42}_{-0.25}$ & 3\\
1990B  & - & 2.64/5.46 & - & - & - & 0.2 & 34\\
1990U  & 15.8$_V$   & 1.6  & 32.64 & 0 $\pm$ 12 & 42.93$^{+0.44}_{-0.20}$ & 0.51$^{+1.91}_{-0.32}$ & 4,5\\
1990W  & 14.8$_V$   & 0.55 & 31.43 & 0 $\pm$ 3  & 42.43$^{+0.60}_{-0.14}$ & 0.16$^{+0.93}_{-0.09}$ & 6,*\\
1990aa & 17.0$_N$ $\pm 0.5$ & 0.175 & 34.13 & 7 $\pm 7$ & 42.50$^{+0.80}_{-0.20}$& 0.19$^{+1.84}_{-0.12}$ & 32,*\\
1990aj & - & -  & - & - & - &0.2 & 33 \\
1991A  & 18.0$_N$  & 1.3   & 33.54 & 1 $\pm$ 10 & 42.19$^{+0.31}_{-0.11}$& 0.094$^{+0.233}_{-0.051}$ & 5,5\\
1991L  & - & -  &  -&- & - & 0.2 & 35 \\
1991N  & 13.9$_N$ & 0.097 & 31.29 & 5 $\pm$ 5 & 42.53$^{+0.56}_{-0.06}$& 0.20$^{+1.06}_{-0.10}$ & 36,*\\
1995bb & - & - &-  & - & - &0.2 & 37\\
1996D  & 18.2$_V$ & 0.509 & 34.04 & 0 $\pm$ 7 & 42.10$^{+0.68}_{0.14}$& 0.07$^{+0.53}_{-0.04}$ & 38,*\\
1996N  & - & - & - &-  &-&0.2  & 29 \\
1996aq & 14.7$_V$  & 0.129 & 32.20 & 0 $\pm$ 4 & 42.61$^{+0.62}_{-0.14}$&0.24$^{+1.48}_{-0.14}$ &$^7,^*$\\
1997B  & 16.5$_N$& 0.243& 33.17& 10 $\pm$ 3 & 42.40$^{+0.52}_{-0.06}$& 0.15$^{+0.70}_{-0.07}$& 39,*\\
1997X  & 13.5$_N$ &0.091 &31.15 & 4 $\pm$ 4 & 42.61$^{+0.54}_{-0.06}$& 0.25$^{+1.21}_{-0.12}$& 40,*\\
2000ew & 14.9$_N$& 0.147 & 30.16 & 14 $\pm$ 7 & 41.88$^{+0.60}_{-0.06}$& 0.05$^{+0.26}_{-0.02}$& 41,*\\
2001ig &   -       &  -    &   -   &     -     & -                      &0.13$^{+0.02}_{-0.02}$ & 8\\
2003bg & 15.0$_N$  & 0.096  & 31.24   & 14 $\pm$ 7 & 42.25$^{+0.60}_{-0.06}$ & 0.11$^{0.61}_{-0.07}$ & 10,*\\
2003dh &   -       &   -   &   -   &     -     &-                       &0.4$^{+0.15}_{-0.1}$ & 9\\
2004ao & 14.9$_N$ & 0.348          & 32.35 & 0 $\pm$ 4 & 42.56$^{0.54}_{-0.06}$& 0.22$^{+1.06}_{-0.11}$ & 13,*\\
2004dk & 17.6$_N$ & 0.522          & 32.15 & $-$10 $\pm$ 10& 41.87$^{+0.66}_{-0.06}$&0.044$^{+0.298}_{-0.021}$ & 15,*\\
2004gk & 13.3$_N$ & 0.10  & 31.02 (Virgo)& 0 $\pm$ 5 & 42.56$^{+0.56}_{-0.06}$ & 0.22$^{+1.13}_{-0.11}$ & 14,*\\
2004gq & 15.5$_N$ &- &- &-4 $\pm$ 3&- & 0.2& 30 \\
2004gt & 14.9$_N$ &  0.22 $\pm$ 0.03 & 31.84 & $-$3 $\pm$ 5 & 42.36$^{+0.17}_{-0.07}$& 0.14$^{+0.21}_{-0.07}$ & 11,12\\
2004gv & 17.6$_N$& 0.110 & 34.50 & 0 $\pm$ 7 &42.38$^{+0.60}_{-0.06}$ & 0.14$^{+0.83}_{-0.07}$ & 27,*\\
2005N  &  - &-  &-  &- &- &0.2 & 42\\
2005bf & -        &     -          &  -    &     -      & - &0.05$^{+0.03}_{-0.03}$ & 26\\    
2005kl & 14.6$_N$ & $>>$ 1  &- & - & - & 0.2 & 22,28\\
2006F  & 16.7$_N$ & 0.629          & 33.70 & 7 $\pm$ 7 & 42.63$^{+0.60}_{-0.06}$& 0.26$^{+1.47}_{-0.12}$ & 19,*\\
2006T  & 17.4$_N$ & 0.246          & 32.68 & $-$11 $\pm$ 2& 42.07$^{+0.58}_{-0.14}$& 0.07$^{+0.383}_{-0.04}$ & 16,*\\
2006gi & 16.3$_N$ & 0.080          & 33.18 & 3 $\pm$ 5  &42.28$^{+0.56}_{0.06}$ &0.11$^{+0.59}_{-0.06}$ & 17,*\\
2006ld & 16.0$_N$ & 0.057          & 33.74 & 9 $\pm$ 4  & 42.73$^{+0.54}_{-0.06}$ &0.33$^{+1.60}_{-0.16}$ & 18,*\\
2007C  & 15.9$_N$ & 0.140          & 32.15 & 2 $\pm$ 4  & 42.03$^{+0.54}_{-0.06}$ &0.065$^{+0.317}_{-0.032}$ & 20,*\\
2007I  & 18.0$_N$ & 0/1.5 & 34.75 $\pm$ 0.25 & 10 $\pm$ 6 & 42.33$^{+0.82}_{-0.10}$ & 0.13$^{+1.33}_{-0.07}$ & 21,*\\
2007bi &  18.3$_N$ &  0/1.5    &  38.8    &  0 $\pm 10$ &43.64$^{+0.86}_{-0.06}$ & 2.6$^{+29.4}_{-1.3}$ & 23,*\\
2007ce & 17.4$_N$ & 0/1.5 & 36.37 &  5 $\pm$ 3 & 43.12$^{+0.72}_{-0.12}$& 0.80$^{+6.33}_{-0.44}$ & 43,*\\
2007rz & 16.9$_N$ & 0.660 & 33.59 & 7 $\pm$ 7 & 42.52$^{+0.60}_{0.06}$& 0.20$^{+1.14}_{-0.10}$ & 24,*\\
2007uy & 16.9$_N$ & 0.075 & 32.48 & $-$7 $\pm$ 7 & 41.84$^{+0.60}_{-0.20}$& 0.041$^{+0.239}_{-0.026}$ & 25,*\\
2008aq & -        & -     &   -   &    -       &  - & 0.2 & 44 \\
\hline
\label{tab2}
\end{tabular}
\caption{Observed magnitude ($V$-band, $B$-band, $N$ = unfiltered
  CCD), extinction (in the $V$ band), distance modulus, epoch,
  estimated peak luminosity, and $^{56}$Ni mass for 41 CC-SNe, together with 
  references for the magnitude and absorption. 
  For SNe 1987M, 2001ig, 2003dh, and 2005bf, estimates for
  the $^{56}$Ni mass {\it or} luminosity are available and referenced. The
  uncertainty on the epoch was taken from \citet{2008ApJ...687L...9M}
  and \citet{2009arXiv0904.4632T}, if available, or
  estimated from the detection report.  SNe without estimate for absorption are
  referenced with ``*.'' For SNe 2007ce, 2007I, and 2007bi, we have no information 
  about the host galaxy, so we assumed an absorption between 0 and 1.5 mag
  (unfiltered). SN 2004gk is a member of the Virgo cluster and
  we therefore took the Virgo distance modulus for this SN, as no
  reliable estimate for the distance of the host galaxy is available. References:
$^1$\citet{1986PAZh...12..784T}, $^2$\citet{1986ApJ...302L..59B}, $^3$\citet{1990BAAS...22.1221N},
$^4$\citet{1997thsu.conf..863C}, $^5$\citet{1994AJ....108..195G}, $^6$\citet{1990IAUC.5080....2E},
 $^7$\citet{1996IAUC.6454....1N}, $^8$\citet{2009arXiv0903.4179S}, $^9$\citet{2005ApJ...624..898D},
$^{10}$\citet{2003IAUC.8082....1W},
$^{11}$\citet{2004IAUC.8454....1M}, $^{12}$\citet{2005ApJ...630L..33M},
$^{13}$\citet{2004IAUC.8299....1S},
$^{14}$\citet{2004IAUC.8446....1Q}, $^{15}$\citet{2004CBET...75....1G},
$^{16}$\citet{2006IAUC.8666....2M},
$^{17}$\citet{2006IAUC.8751....2I}, $^{18}$\citet{2006IAUC.8766....1F},
$^{19}$\citet{2006CBET..364....1D},
$^{20}$\citet{2007IAUC.8792....2P}, $^{21}$\citet{2007IAUC.8798....1J},
$^{22}$\citet{2005CBET..300....1D},
$^{23}$\citet{2007CBET..929....1N}, $^{24}$\citet{2007CBET.1158....1P},
$^{25}$\citet{2008CBET.1191....2B},
$^{26}$\citet{2007ApJ...666.1069M}, $^{27}$\citet{2004IAUC.8454....1M},
$^{28}$\citet{2005CBET..305....1T} 
but absorption highly uncertain, assume a $^{56}$Ni mass of 0.2 M$_\odot$, 
$^{29}$\citet{1996IAUC.6351....1W} but no light-curve
 information, assume a $^{56}$Ni mass of 0.2 M$_\odot$, 
$^{30}$\citet{2004IAUC.8452....2P} bu no distance information available,
assume a $^{56}$Ni mass of 0.2 M$_\odot$, 
$^{31}$\citet{1996ApJ...459..547C}, $^{32}$\citet{1990IAUC.5087....1P}, 
$^{33}$\citet{1991IAUC.5178....1M} but no light-curve information available, 
assume a $^{56}$Ni mass of 0.2 M$_\odot$,
$^{34}$\citet{2001ApJ...553..886C} but distance and reddening highly uncertain,
assume a $^{56}$Ni mass of 0.2 M$_\odot$, $^{35}$\citet{1991IAUC.5200....1P} but no 
light-curve information, assume a $^{56}$Ni mass of 0.2 M$_\odot$,
$^{36}$\citet{1991IAUC.5234....1F}, $^{37}$\citet{1995IAUC.6271....1T} but no
light-curve information available, assume a $^{56}$Ni mass of 0.2 M$_\odot$,
$^{38}$\citet{1996IAUC.6317....2D}, $^{39}$\citet{1997IAUC.6535....1G},
$^{40}$\citet{1997IAUC.6552....1N}, $^{41}$\citet{2000IAUC.7530....1P}, 
$^{42}$\citet{2005IAUC.8472....2S} but no light-curve information, assume a 
$^{56}$Ni mass of 0.2 M$_\odot$, $^{43}$\citet{2007CBET..953....1Q}, 
$^{44}$\citet{2008ATel.1403....1B} but no light-curve information and no 
detailed epoch information, assume a $^{56}$Ni mass of 0.2 M$_\odot$.}
\end{table*}

\begin{table*}
\begin{tabular}{|l|c|c|c|c|c|c|c|}
\hline
SN   & Type & Epoch (days) & 1D & $v_{\alpha}$ (km s$^{-1}$) & $v_{50}$ (km s$^{-1}$) & Ref.\\
\hline
1983N  & Ib   & 226  & Y  & 3797& 2630& $^*$\\
1985F  & Ib/c & 280  &  Y & 4920& 2456&$^*$\\
1987M  & Ic   & 141  &  Y & 5486& 3701& $^*$\\
1990B  & Ic   & 140 & N   & 5405& 5091&$^*$\\
1990I  & Ib   & 237  &  Y & 4828& 2899& $^*$\\
1990U  & Ic   & 184  &  Y & 3488 & 3021&$^*$ \\
1990W  & Ib/c & 183  &  Y & 4803 &3425& $^*$\\
1990aa & Ic   & 141  & ?  & 4216& 4368&$^*$\\
1990aj & Ib/c & 150-250&? & 5122& 3034&$^*$\\
1991A  & Ic   & 177  &  Y & 5262 &3636& $^*$\\
1991L  & Ib/c & 100-150&N & 4140 & 3316&$^*$\\
1991N  & Ic   & 274  &  Y & 4278 & 3239&$^*$\\
1993J  & IIb  & 205  &  Y & 4029 &3070&$^*$ \\
1994I  & Ic   & 147  &  Y & 5057 &3967&$^*$ \\
1995bb & Ib/c & 150-400&Y & 5154 & 4410&$^*$\\
1996D  & Ic   & 214  &  Y & 5228 & 3624&$^*$\\
1996N  & Ib   & 224  &  N & 3736 & 3047&$^*$\\
1996aq & Ib   & 226  &  N & 5846 &3451& $^*$\\
1997B  & Ic   & 262  &  Y & 4801& 3317&$^*$\\
1997X  & Ic   & 103  &  Y & 4680 &3420& $^*$\\
1997dq & BL-Ic& 217  &  Y & 4594 &3361&$^*$ \\
1997ef & BL-Ic& 287  &  Y & 4681 &2733&$^*$ \\
1998bw & BL-Ic& 201  &  Y & 6340 &3602&$^*$ \\
2000ew & Ic   & 122  &  N & 4184 &3001&$^*$\\
2001ig & IIb  & 256  &  N & 5027 &3241& \citet{2009PASP..121..689S}\\
2002ap & BL-Ic& 185  &  Y & 6219 &3729&$^*$ \\
2003bg & BL-IIb& 279 &  N & 4736 &3205& Hamuy et. al. (in prep.)\\
2003dh & BL-Ic& 229  &  ? & 4342 &3085& Bersier et. al. (in prep.)\\
2003jd & BL-Ic& 317  &  N & 6850 &5593&$^*$ \\
2004ao & Ib   & 191  &  N & 4555 &3158& \citet{2008ApJ...687L...9M}\\
2004aw & Ic   & 236  &  Y & 5007 &3234& $^*$\\
2004dk & ?    & 333  &  Y & 5465 & 4338&\citet{2008ApJ...687L...9M} \\
2004gk & Ic   & 225  &  Y & 4623 &3171& \citet{2008ApJ...687L...9M}\\
2004gq & Ib   & 297  &  Y & 6697 &3039& \citet{2008ApJ...687L...9M} \\
2004gt & Ic   & 160  &  N & 4513 &3371& $^*$\\
2004gv & Ib/c & 299  &  Y & 4792 &3158& \citet{2008ApJ...687L...9M} \\
2005N  & Ib/c &70-120&  N & 4001 & 3591&$^*$\\
2005bf & Ib   & 209  &  N & 3864 & 3628&\citet{2008ApJ...687L...9M} \\
2005kl & Ic   & 160  &  Y & 5074 & 3365&\citet{2008ApJ...687L...9M} \\
2006F  &  Ib  & 314  &  N & 4491 & 2966& Mazzali\\
2006T  &  ?   & 371  &  N & 4202 & 4436& $^*$\\
2006aj & BL-Ic& 204  &  Y & 6540 & 5100&$^*$\\
2006gi &  ?   & 148  &  Y & 4589 & 3180& $^*$\\
2006ld &  Ib  & 280  &  N & 4086 & 3182& $^*$\\
2007C  & Ib   & 165  &  N & 4787 & 3520&$^*$ \\
2007I  & BL-Ic& 165  &  N & 6085 & 3277& $^*$\\
2007Y  &  Ib  & 270  &  Y & 4331 & 3025& \citet{2009ApJ...696..713S}\\
2007bi & PI?  & 360  &  Y & 5487 & 3756 & Mazzali\\
2007ce & BL-Ic& 310  &  Y & 6172 & 4461& Matheson\\
2007gr & Ic   & 158  &  Y & 4480 & 3228& \citet{2009Natur.459..674V}\\
2007ru & BL-Ic& 200  & Y  & 5811 & 3981& \citet{2009ApJ...697..676S}\\
2007rz & Ic   & 292  &  ? & 4998 & 3785& Mazzali\\
2007uy & Ib   & 111  &  Y & 6103 & 4563& Mazzali\\
2008D  & BL-Ib&  86  &  N & 5847 & 4040 & Mazzali \\
2008aq & IIb  &$>$130&  Y & 4119 & 2885& Matheson\\ 
2008ax &  IIb & 246  &  N & 4100 & 2821& Navasardyan et. al. (in prep.)\\
\hline
\label{tab3}
\end{tabular}
\caption{SN type (classified by early-phase spectra), the epoch of
our spectra (relative to maximum light), and the characteristic velocities
$v_\alpha$ and $v_{50}$ both corrected by 5\% per 100 days. The
maximum errors of $v_{\alpha}$ and $v_{50}$ are
estimated to be $\pm$ 14\% and $\pm$10\%, respectively. SNe marked with ``Y'' are 
fit by our 1D shell modelling quite well. For SNe marked with ``N'' the central
parts of the line profiles are not reproduced well and therefore probably 
could be improved with multi-dimensional
modelling ($v_\alpha$ would not change much, as $v_\alpha$ is dominated by the
outer parts of the line profiles, where most of the kinetic energy is located). 
For SNe marked with ``?'' the spectra are too noisy to
categorise as ``Y'' or ``N.'' In total we have 12 BL-SNe (2 of them are
GRB hypernovae), 17 regular SNe~Ic (27 with BL-Ic), 12 regular SNe~Ib (13 with
BL-Ib), 4 regular SNe~IIb (5 with BL-IIb), 1 possible
PI SN, and 10 SNe for which the classification is unclear (they are
certainly of the CC-SN type). See \citet{2009arXiv0904.4632T} for spectra referenced
by ``*.''}
\end{table*}

\section{Spectral Modelling}

The spectra were modelled using the nebular code of
\citet{2001ApJ...559.1047M,2007ApJ...670..592M} in the stratified version. A
Monte Carlo routine is used to calculate the deposition of energy (which is
carried by the $\gamma$-rays and positrons produced by the decay of $^{56}$Ni
and $^{56}$Co) in each shell. The gas heating caused by this process is balanced
by cooling via line emission. The excitation and ionisation state of the gas,
together with the electron density and temperatures, are then iterated until the
line emissivity in each shell balances the deposited energy. The emission rate
in each line is calculated solving a non-LTE matrix of rates to
obtain the level populations
\citep{1980PhDT.........1A}.

The general principle of such modelling is that different velocity shells are
characterised by different element abundances and densities, which lead to
different ratios of line fluxes in each shell. The total emissivity of a shell
is controlled by the energy deposition, which depends on the density and the
$^{56}$Ni mass of the shell and the neighbouring ones. Therefore, each shell has
a certain emissivity integrated over all wavelengths and a characteristic line
profile caused by the Doppler shift. Each line in the emerging spectrum is the
result of the superposition of single components from each shell, with different
characteristic widths. A change of the emissivity of one line results in a
variety of profile variations of other lines which have to be modelled
iteratively until the complex shape of the full spectrum is reproduced. As a
complicated structure must be found in order to produce a certain spectrum, the
method is quite reliable in determining the velocity field of the ejecta.

We start modelling each spectrum using a CC-SN model used by
\citet{2002ApJ...572L..61M} for SN 2002ap. Such a model contains all of 
the information about density, mass, velocity, and element abundances.
We correct the $^{56}$Ni mass to the values listed in Tables 1 and 2,
and we scale the synthetic spectrum to match the observed one.

A clear modelling of the Fe-group lines is impossible for most nebular spectra
because these lines are generally weak and are therefore affected by noise and
background. The oxygen line is therefore taken as a tracer for the $^{56}$Ni
distribution. This means that our $^{56}$Ni zone extends out to the point where
the oxygen line can no longer be separated clearly from the background.
Hence, a narrow oxygen line results in a more central $^{56}$Ni distribution
in our models, but it is important to note that this assumption may not
exactly reflect the situation in real SNe. This may cause some epoch-dependent
error which we try to quantify by comparing the time dependence of line widths in
our models and the observed line widths (see Section 4).  Above the $^{56}$Ni/O 
zone we set the density to zero, as this region cannot be probed with
the nebular approach. We discuss this in more detail in Section 4.

Other elements, such as O, Ca, and C, are distributed over the velocity shells
until the line profiles are matched. We pay special attention to fitting
the O line, since oxygen is typically the most abundant element in
stripped-envelope CC-SNe.

\begin{figure}   
\begin{center} 
\includegraphics[width=8.5cm,clip]{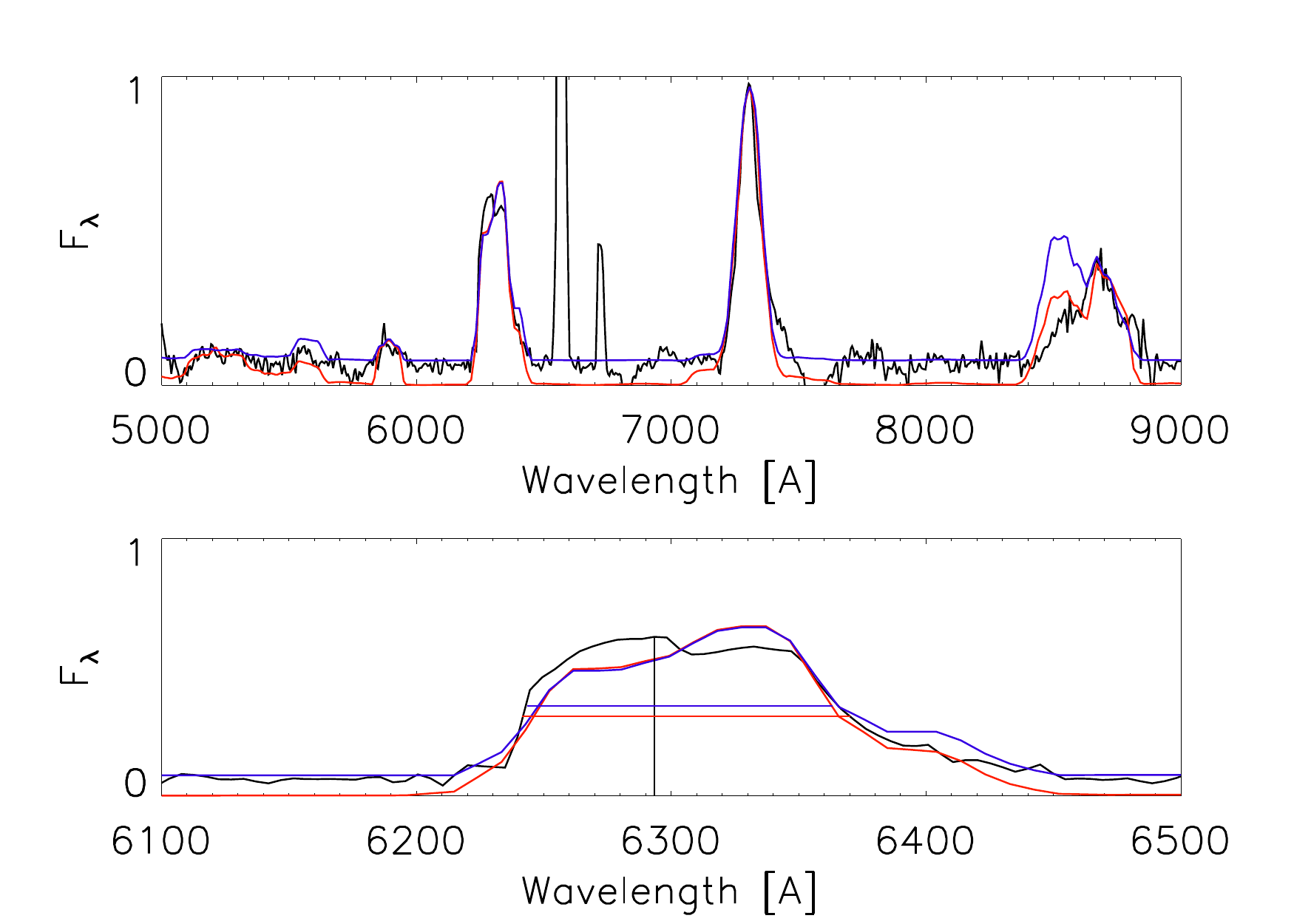}
\end{center}  
\caption{Nebular spectrum of SN 1990U ($v_\alpha =
3450$(blue)/3633(red) km s$^{-1}$) at $t = 184$ days (black) (upper panel) and the
oxygen doublet [O {\sc i}] $\lambda\lambda$6300, 6364 isolated
(lower panel). Two  models are shown in blue and red, illustrating the
uncertainty in the background. This spectrum is the one with the lowest $v_\alpha$ in
our sample (narrow oxygen line). The characteristic velocity is $v_{50}$ =
2996(blue)/3124(red) km s$^{-1}$. The horizontal lines show the full width
at half-maximum intensity (FWHM).}
\label{fig2} 
\end{figure}

\begin{figure} 
\begin{center}
\includegraphics[width=8.5cm, clip]{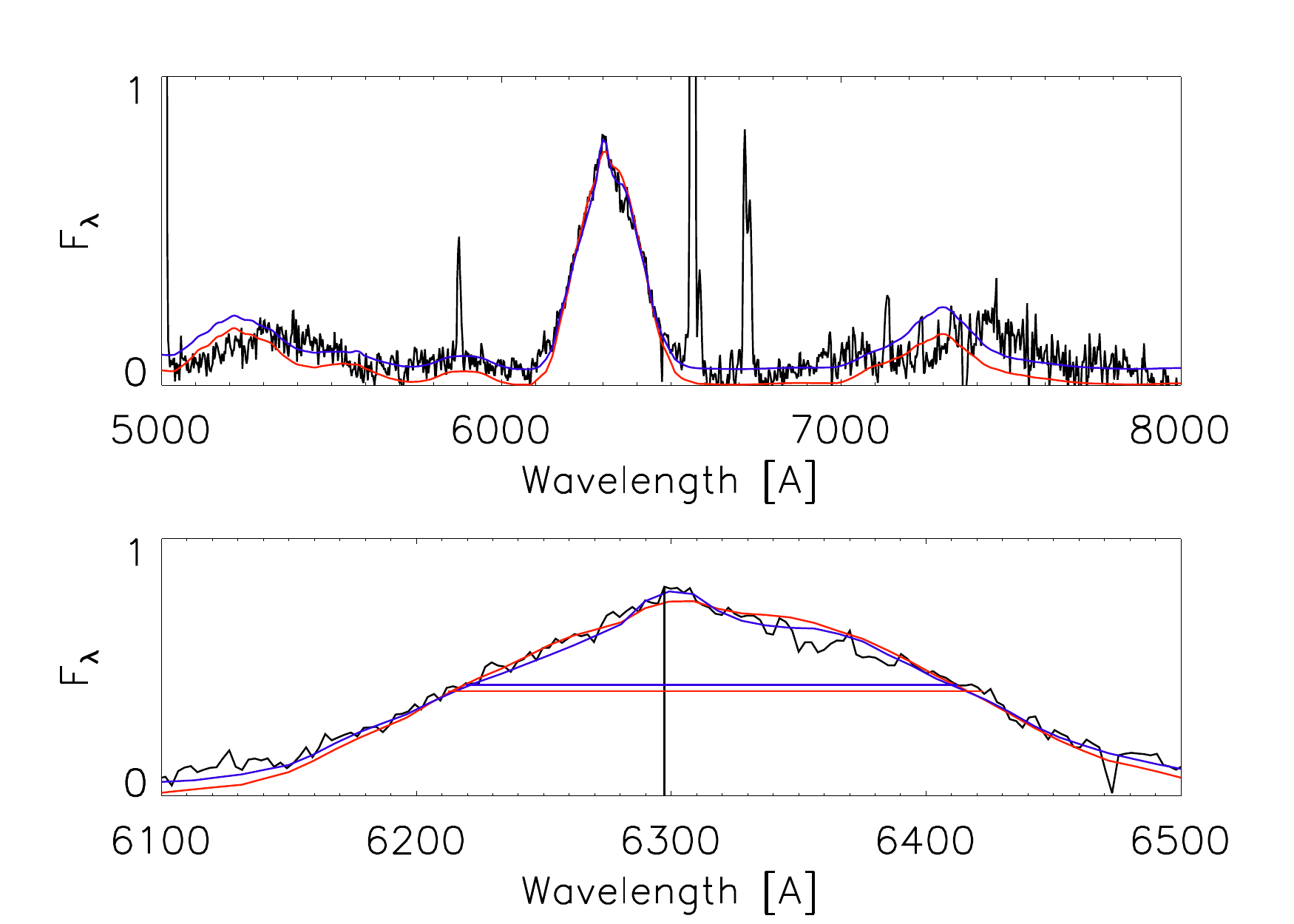}
\end{center}
\caption{Nebular spectrum of the BL-SN 2006aj ($v_\alpha
  = 6542$(blue)/6870(red) km s$^{-1}$) at $t = 204$ days (black) (upper
  panel) and the oxygen doublet [O {\sc i}] $\lambda\lambda$6300, 6364 isolated
  (lower panel). Two  models are shown in blue and red, illustrating the
  uncertainty in the background. This spectrum has one of the
  highest $v_\alpha$ in our sample (broad oxygen line). The
characteristic velocity $v_{50}$ = 4992(blue)/5190(red) km s$^{-1}$. The
horizontal lines show the FWHM.}
\label{fig3}
\end{figure}

The modelling process yields the abundances, masses, and velocities which
best reproduce the observed spectra (see Figures \ref{fig2} and \ref{fig3} for
two examples). The mass and velocity distribution can then be integrated to
obtain the total mass and kinetic energy of the model. The ratio of core
kinetic energy (in units of 10$^{50}$ ergs) to core mass (in units of M$_\odot$) 
is termed $\alpha$ in this paper: 
\begin{equation} 
\alpha \equiv \frac{E_{\rm kin,50}}{M_{\rm ej}({\rm M}_\odot)}. 
\end{equation} 
This parameter is measured for all 56 CC-SNe of our sample and is converted to 
a characteristic velocity (see Table 3),

\begin{equation}
v_{\alpha} = \Big(\frac{2\alpha
  10^{40}}{{\rm M}_{\odot}{\rm [g]}}\Big)^{1/2} {\rm [km~s^{-1}]}.
\end{equation}

The largest uncertainty in the estimate of $\alpha$ is caused by the background.
The kinetic energy is dominated by the outer parts of the oxygen line and these
are superposed on background lines and noise, which are difficult to
distinguish. To quantify the uncertainty that this causes, we tried different
fits for several spectra (see Figure \ref{fig2} and \ref{fig3} for two
examples). We found that, depending on the background assumed, variations of up
to $\pm 15$\% in $\alpha$ can occur, which translates into an uncertainty of
$\sim 7$\% in $v_{\alpha}$. An exact treatment of this problem is difficult, as
background subtraction is arbitrary to some degree. Therefore, based on
this uncertainty, we estimate an error of 7\% in $v_\alpha$.

As we show in Section 4, the line width slowly evolves with time, causing
$v_{\alpha}$ to decrease by roughly 5\% every 100 days. To handle this problem,
we corrected $v_{\alpha}$ by $(1 + 0.05(t{\rm (days)} - 200)/100)$ and assume that
this causes an additional error of $\pm 5$\% in $v_{\alpha}$.

\begin{figure} 
\begin{center}
\includegraphics[width=8.5cm, clip]{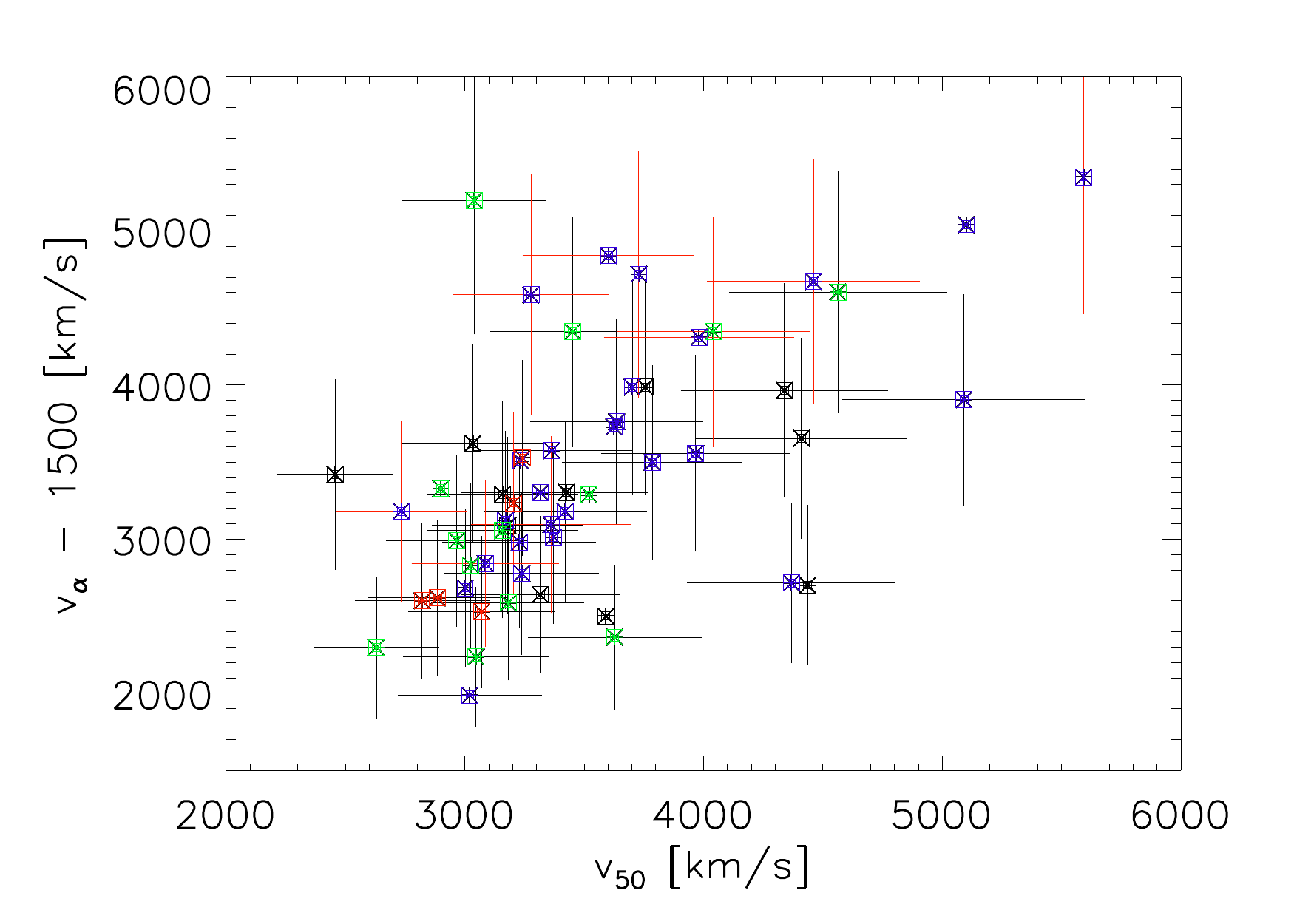}
\end{center}
\caption{The characteristic velocities $v_{50}$ [km s$^{-1}$] vs. ($v_{\alpha}$ $-$
1500 km s$^{-1}$). Both characteristic velocities seem to correlate (as
expected). The offset by $\sim$1500 km s$^{-1}$ is caused by the different
velocities cutoffs used to calculate $v_{50}$ (half height) and $v_{\alpha}$
(full line profile). BL-SNe are shown in red. SNe~Ic are marked with
blue, SNe~Ib with green, and SNe~IIb with red dots. Even taking 
into account the substantial errors, it is obvious that BL-SNe have a large scatter 
of characteristic core velocities. The data point at the upper right is SN 2003jd
(see Table 3). The four BL-SNe at low characteristic velocities are SNe 1997ef,
1997dq, 2003dh, and 2003bg. The intermediate characteristic velocity BL-SNe are
(from the left to the right) SNe 2007I, 1998bw, 2002ap, 2008D, 2007ce, and 2006aj.
The three BL-SNe 2007I, 1998bw, and 2002ap fall off the $v_{\alpha}$/$v_{50}$
relation owing to the strong broadening toward the base of their [O {\sc i}]
$\lambda\lambda$6300, 6364 line.}
\label{fig5}
\end{figure}

In addition to $v_\alpha$, we also measure the half width of the oxygen doublet
[O {\sc i}] $\lambda\lambda$6300, 6364 at half-maximum intensity (HWHM),
$\Delta_{50}$, for all SNe of our sample. This can be directly converted to a
velocity via the redshift

\begin{equation}
v_{50} = \frac{\Delta_{50}}{6300}c,
\end{equation}

where $c$ is the speed of light and $\Delta_{50}$ is given in \AA\ units. 
This is another characteristic velocity of
the SN core ejecta, which can be compared to $v_\alpha$ (see Figure
\ref{fig5}). There seems to be a linear trend between $v_{50}$ and $v_{\alpha}$,
as expected; exceptions are mostly caused by specific features of some
line profiles, as discussed in detail in Section 4. 

Estimating $v_{50}$ has
the advantage that it can be done easily and it does not require modelling. The
disadvantage is that $v_{50}$ contains some small contribution of the [O {\sc
i}] $\lambda$6364 line which causes some error. Since the ratio of [O {\sc i}] 
$\lambda$6300 to [O {\sc i}] $\lambda$6364 is $\sim$3 in the nebular phase, this
error is small; it depends on the exact shape of the individual lines, but it
should be $<$20\% in the worst case, as we found by superposing
Gaussians. More importantly, it is influenced by the shape of the inner line
profile (which in some SNe has a double- or even triple-peaked shape) much more
than by $v_{\alpha}$. In addition, $v_{50}$ measures the velocity at one point
(half height), while $v_\alpha$ is the result of an integration over the entire
line profile and includes the nonlinear weighting of different emission
regions. The background causes some uncertainty in the estimate of
$v_{50}$. We measure $v_{50}$ for several SNe while varying our assumptions
about the background, and estimate an uncertainty of $\sim$5\%. Moreover, we
apply the same time-dependent correction as for $v_{\alpha}$ and assume an
additional error as for $v_\alpha$ ($\pm 5$\%). Adding the possible errors due
to background, $^{56}$Ni mass estimate, and time evolution of the line width, we
estimate a maximum error of $\pm 14$\% for $v_\alpha$ and $\pm 10$\% for
$v_{50}$, which does not necessarily mean that $v_{50}$ gives a more correct 
estimate of the core velocity.

\section{Discussion}
\subsection{Tests}
\label{ism} 

\begin{figure} 
\begin{center}
\includegraphics[width=8.5cm, clip]{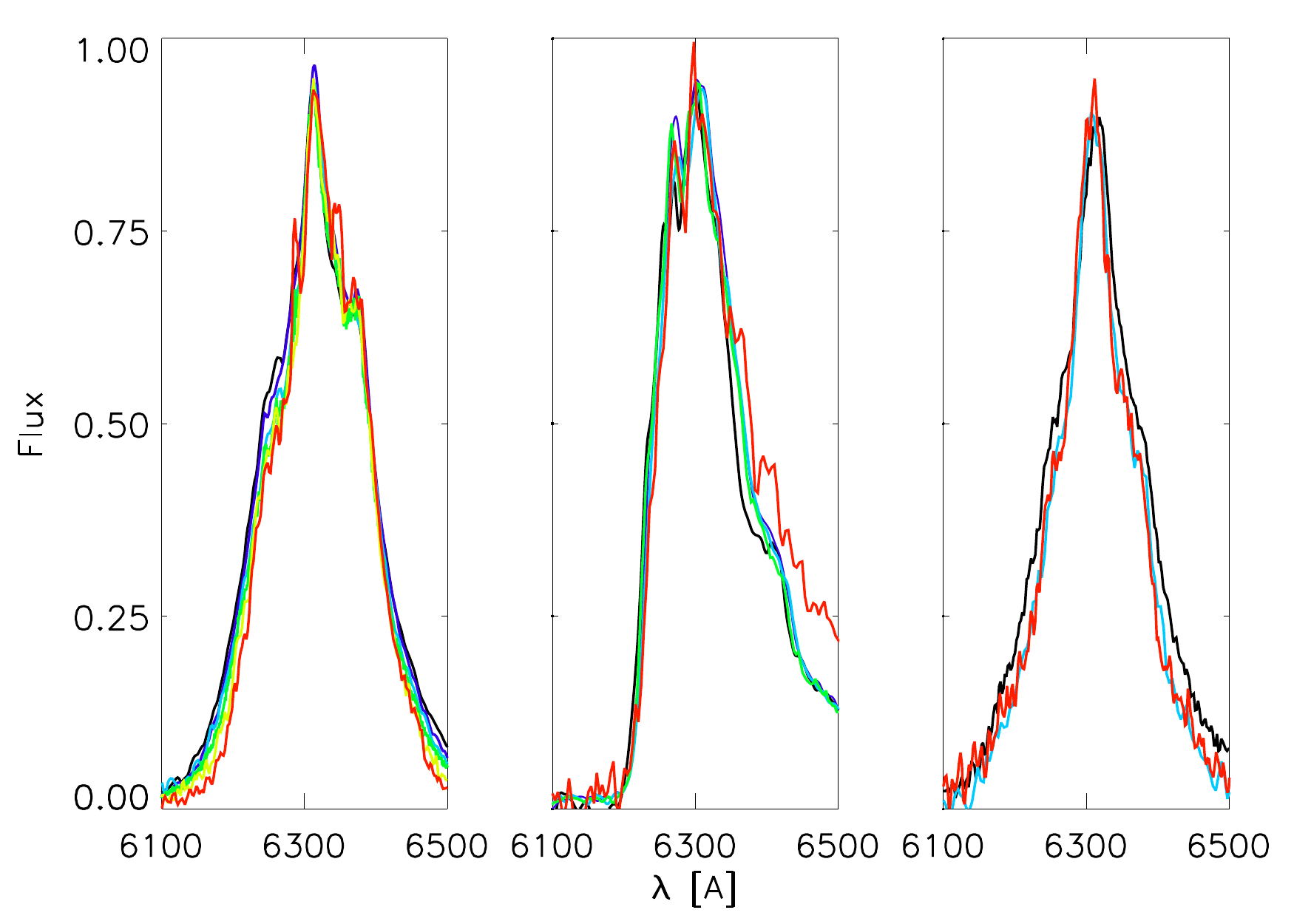}         
\end{center}
\caption{The [O {\sc i}] $\lambda\lambda$6300, 6364 line profiles (left
to right) of SN 2002ap (day 156, black; day 185, blue; day 237, violet; day
274, green; day 336, yellow; day 386, red), SN 1993J (day 205, black; day 
236, blue; day 255, green; day 295, yellow; day 367, red), and SN 1998bw 
(day 201, black; day 337, red; day 376, blue). The temporal
evolution is quantified in Figure \ref{fig4}.}
\label{fig10}
\end{figure}

\begin{figure} 
\begin{center}
\includegraphics[width=8.5cm, clip]{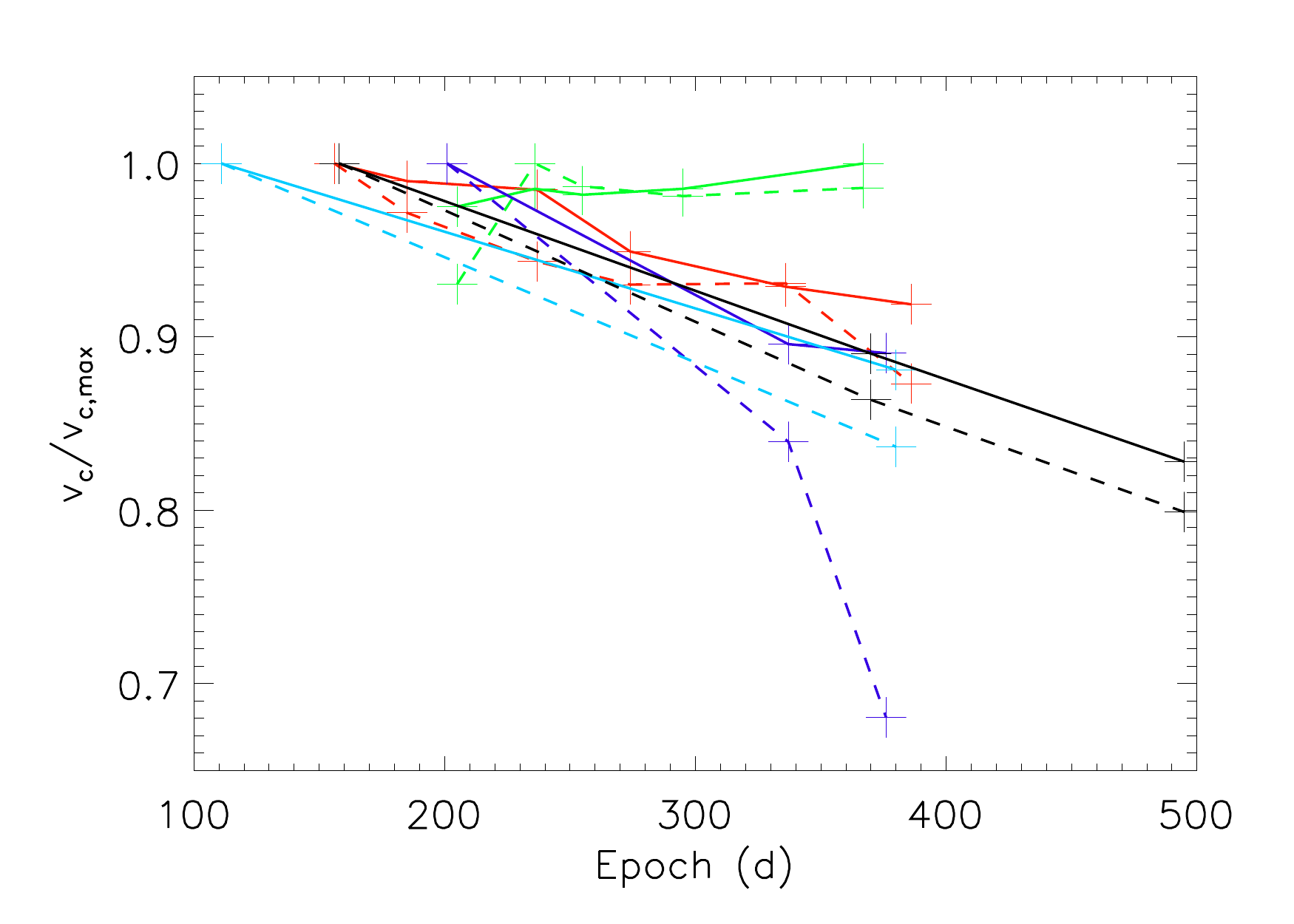}         
\end{center}
\caption{Temporal evolution of $v_\alpha$ (full line) and $v_{50}$
  (dashed line) for SNe 2002ap (red), 1993J(green), 1998bw (blue),
  2007uy (light blue), and 2007gr (black). One can
see that variations are rather small, about 5\% per 100 days. The
rapid drop of $v_{50}$ of SN 1998bw is caused by the ``knot'' at
half height of the oxygen-line profile, which ``drops below'' half height
in the late phase. The increase of line width with time in SN 1993J must
be some background effect (note the rise of the red wing at late
times, due to H$\alpha$ emission) which could not be subtracted. For
SNe 1993J, 1998bw, and 2002ap, similar temporal evolution was found 
by \citet{2009arXiv0904.4632T}.}
\label{fig4}
\end{figure}

\begin{figure} 
\begin{center}
\includegraphics[width=8.5cm, clip]{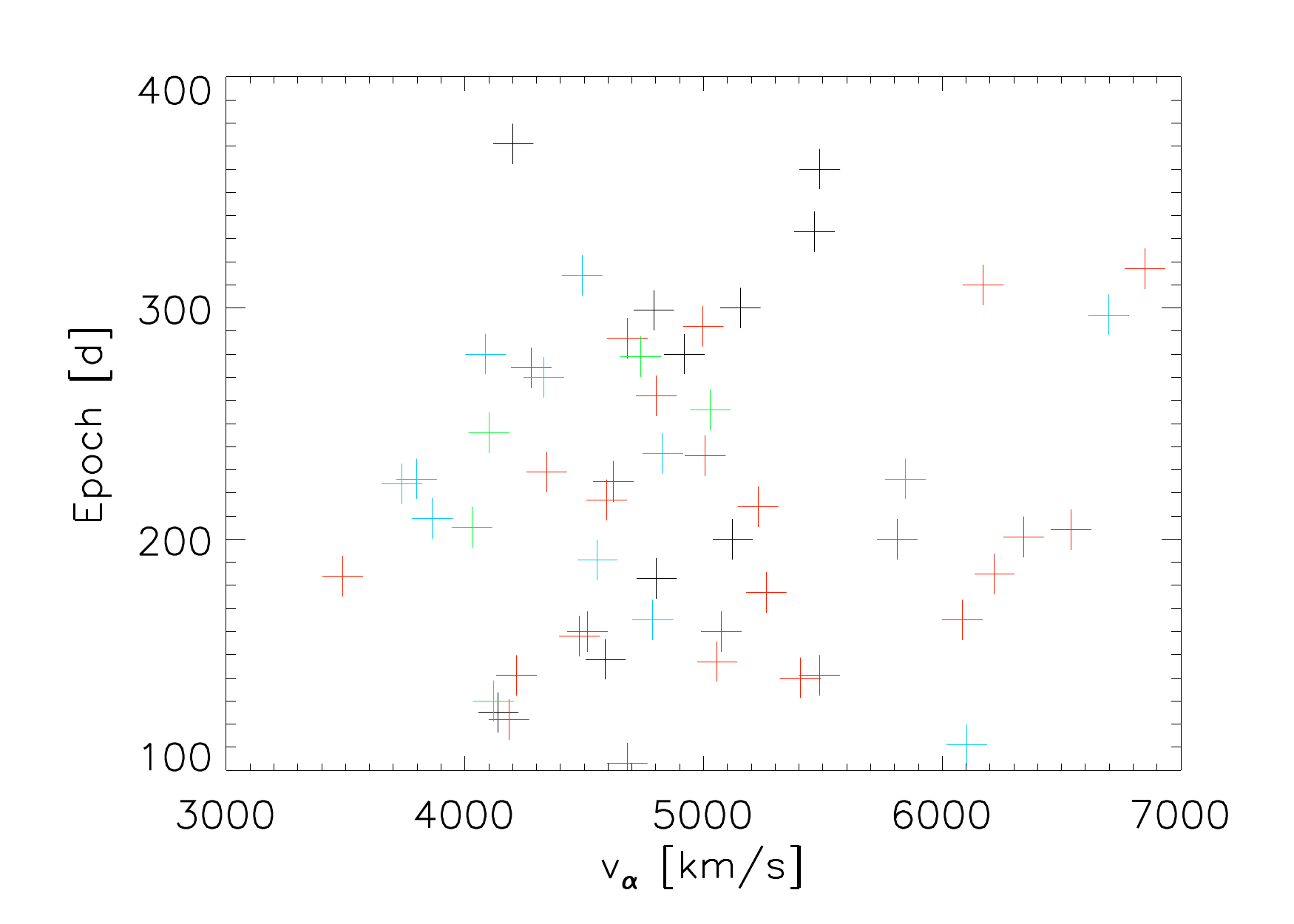}         
\end{center}
\caption{The epoch (days after maximum light) vs. characteristic
  velocity $v_{\alpha}$ (km s$^{-1}$) for all SNe at epochs listed in
  Table 3. SNe~Ic are shown in red, SNe~Ib in blue, and SNe~IIb in
  green. SNe without type classification are shown in black.
  The effect of temporal line-width evolution is weak.}
\label{fig8}
\end{figure}

In order to ensure that the estimates of $v_\alpha$ are reliable, it is useful
to verify that the data do not show behaviours that are not included in our
modelling.  Since the spectra that we used were obtained at very different SN
epochs, it is vital that $v_\alpha$ does not depend strongly on SN epoch. 
Figure \ref{fig10} shows the time evolution of $v_\alpha$ for three SNe with 
high-quality nebular spectra at several different epochs. The temporal evolution 
of $v_\alpha$ is in fact weak, at most $\sim$10\% over 200 days, which is
comparable to the general modelling uncertainty. Figure \ref{fig4} shows the
corresponding evolution of the line profiles.  In Figure \ref{fig8}, $v_\alpha$
is plotted against SN epoch for the entire sample; no strong time dependence is
seen. For $v_{50}$ the situation is similar. However, as can be seen for example
in SN 1998bw (see Figures \ref{fig10} and \ref{fig4}), small features in the
line profile can have disturbing effects on $v_{50}$. The influence of epoch on
the characteristic velocities will be more fully discussed in Section 4.2.

\begin{figure} 
\begin{center}
\includegraphics[width=8cm, clip]{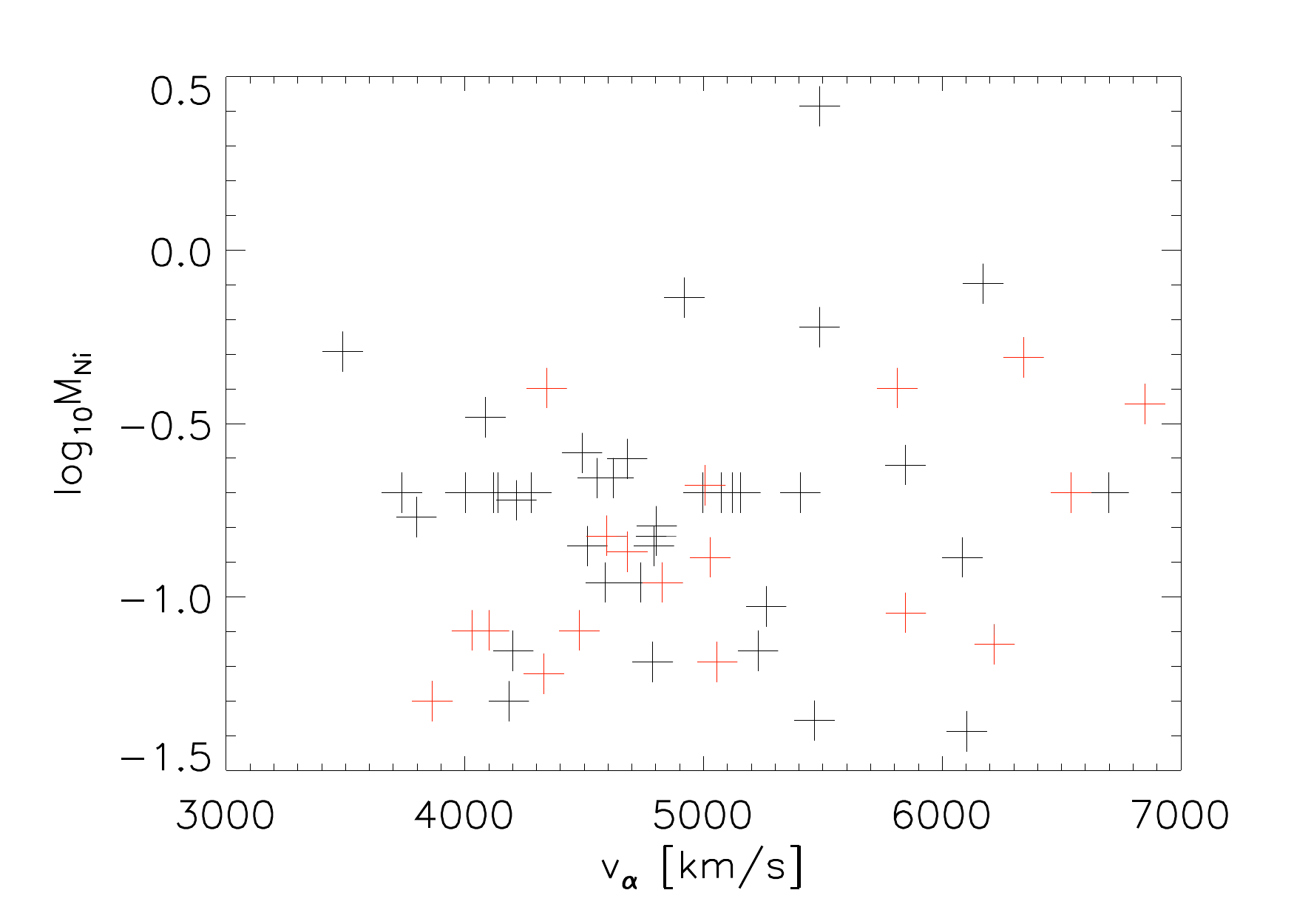}         
\end{center}
\caption{The logarithm of $^{56}$Ni mass (in solar mass units) against the 
  characteristic velocity
  $v_{\alpha}$. No relation is seen for our full sample. There might
  be a weak correlation for the SNe with $^{56}$Ni mass estimates taken from the
  literature (shown in red), where the errors in the $^{56}$Ni mass are much
  smaller. The mean $^{56}$Ni mass of observed CC-SNe seems to lie around
  0.20 M$_\odot$ (0.176 M$_\odot$ for $^{56}$Ni mass estimates taken from
  the literature, 0.225 M$_\odot$ for the $^{56}$Ni mass estimates in this
  paper), with a broad diversity. Neglecting the possible PI-SN 2007bi,
  $^{56}$Ni masses range from 0.05 M$_{\odot}$ to 0.80
  M$_{\odot}$. Because of
the large uncertainties of our estimates, these upper and lower limits
are inaccurate.}
\label{fig9}
\end{figure}

To test whether the often highly uncertain estimate of the $^{56}$Ni mass
influences $v_\alpha$, we alternatively increased and reduced the $^{56}$Ni mass
by a factor of five, therefore spanning a factor of 25 in $^{56}$Ni mass for
several randomly selected SNe of our sample. This corresponds roughly to the
maximum uncertainty in the $^{56}$Ni mass estimates. We found that $v_\alpha$ 
depends only weakly on $^{56}$Ni mass: the difference is always less than 2\%.
For example, when changing the $^{56}$Ni mass of SN 1987M by a factor of 25, from 
0.6 to 0.024 $M_\odot$, $v_\alpha$ increased by only 1.3\%, which is considerably
smaller than other modelling uncertainties. Figure \ref{fig9} shows $v_\alpha$
against $^{56}$Ni mass; there is clearly no correlation. We conclude that the
maximum error introduced by the uncertainty in the $^{56}$Ni mass is $\pm 2\%$.
The independence of $v_\alpha$ on $^{56}$Ni mass is discussed in more detail
in Section 4.2.

\subsection{Discussion of the Method}
\label{disc}

We obtained characteristic velocities for 56 CC-SNe. Before we turn to the
results we discuss our methodology.

In Section 2 we tried to determine the $^{56}$Ni mass for 29 SNe in our sample
without calibrated spectra. A typical uncertainty in $^{56}$Ni mass is a
factor of 5--10 up and a factor of 2--3 down. The uncertainty upward is mainly
caused by the uncertainty in absorption and epoch. There are also uncertainties
in distance, magnitude, converting magnitude to bolometric luminosity and
finally to $^{56}$Ni mass, which are smaller. Our $^{56}$Ni masses might be
systematically over- or underestimated and one should be cautious when using
these values. The $^{56}$Ni masses listed in the literature (18 SNe) have an
average of 0.176 M$_\odot$, while the SNe with $^{56}$Ni masses estimated in this
paper have an average $^{56}$Ni mass of 0.225 (28 SNe, excluding 9 SNe where we
could not estimate a $^{56}$Ni mass and excluding a possible PI-SN).
Considering the relatively small numbers, the agreement is reasonable
but might indicate a small systematic over-estimate of $^{56}$Ni mass.
In any case, as shown in Section 4.1, an uncertainty of a factor of 25 in 
$^{56}$Ni mass will not influence our results significantly.

We checked for possible correlations between $^{56}$Ni mass and $v_\alpha$. This
could be important in two different ways. First, the uncertain estimate of the
$^{56}$Ni mass might introduce some error in the determination of $v_\alpha$.
This is not the case, however, as shown in Section 4.1. Increasing the $^{56}$Ni
mass by a constant factor throughout the ejecta will increase the energy
deposited in any shell by the same factor; hence, the relative contribution
of each shell to the emission will remain constant. As the emitted energy per 
particle also remains rather constant (the mass of other elements must be
increased accordingly to the $^{56}$Ni mass), the spectrum of each
shell does not change much. Therefore. the emerging spectrum is nearly constant
apart from small differences arising from the small shift of line ratios in
each shell (resulting, for example, from changes of the ionisation balance).

Second, there could be a physical correlation between $^{56}$Ni mass and core
ejecta velocities. For our full sample, this does not seem to be the case, as
shown in Figure \ref{fig9}. However, our large uncertainties in the $^{56}$Ni mass 
make a stringent conclusion impossible. There might be a weak correlation for the
subgroup of 18 CC-SNe with $^{56}$Ni masses taken from the literature (the 
uncertainties in these estimates should be much smaller), but the situation is 
not entirely clear owing to the substantial scatter.

\begin{figure} 
\begin{center}
\includegraphics[width=8.5cm, clip]{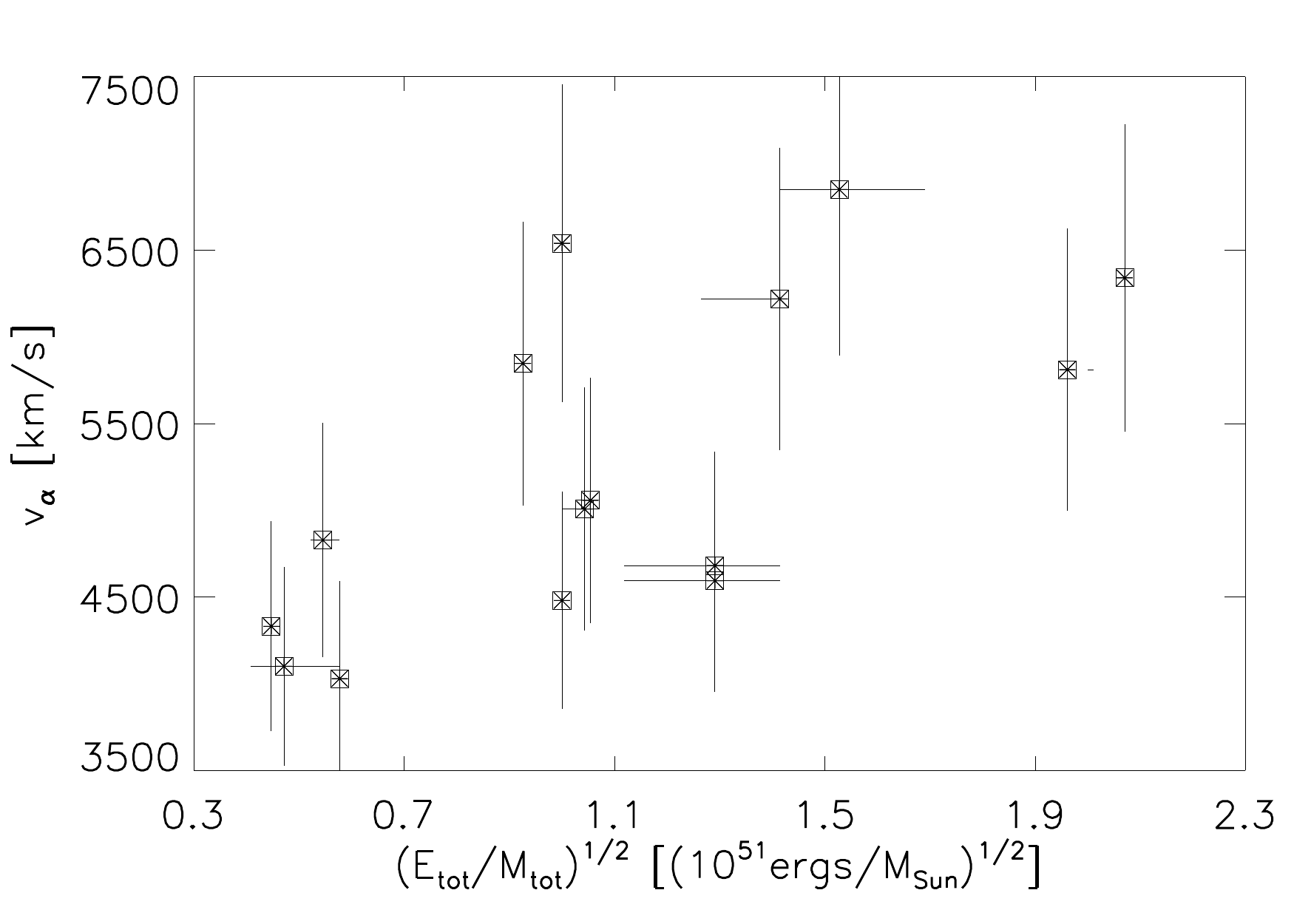}         
\end{center}
\caption{The characteristic velocity $v_\alpha$ [km s$^{-1}$] against the ratio
  $(E_{\rm ej,tot}/M_{\rm ej,tot})^{1/2}$
  [($10^{51}$ ergs/M$_\odot$)$^{1/2}$], which is a proxy for the outer
  ejecta velocity. There seems to be a weak correlation; however,
the scatter is large and there is almost no predictive power in that relation.}
\label{fig666}
\end{figure}

Given that in most spectra we cannot determine the $^{56}$Ni velocity very 
accurately (in CC-SN nebular spectra Fe-group lines are usually weak, 
strongly overlapping, and cannot be easily separated from the background), we 
can draw no conclusion
about possible relations between $^{56}$Ni mass and Fe-group element velocities,
which might differ from light-element velocities in aspherical SN models.
However, there is no correlation between the total kinetic energy or the total
ejecta mass and the characteristic core velocity (for the 15 SNe listed in Table
1). A weak correlation of the ratio $(E_{\rm ej,tot}/M_{\rm ej,tot})^{1/2}$ with 
the characteristic core velocity is found, albeit with large scatter
(see Figure \ref{fig666}).

The epoch at which a SN reaches its nebular phase generally depends on the mass
and the ejecta velocity. Some of our spectra are quite early ($\sim$90 days
after maximum light), though they all seem to be sufficiently nebular to be
treated with our modelling. The epochs of our spectra vary between 100 and 400
days (after maximum light), and we have shown in Section 4 that this large span
of epochs will not influence our results much. From detailed modelling of five
SNe with good spectral coverage in the nebular phase (extending over 200 days),
we can estimate that the characteristic velocity may decrease by $\sim$5\% 
every 100 days. This is caused by the decreasing importance of the outer
layers, which become less luminous. 

In principle, if the behaviour of the
positrons and the detailed distribution of $^{56}$Ni were known, we could
reproduce this line-width evolution. However, since these parameters are
unknown, it is necessary to make some assumptions about both properties. As
described in Section 3, we assumed that both the $^{56}$Ni distribution and the
positron deposition trace the oxygen line profile, leading to an oxygen line
with constant width (since we assume local deposition of positron energy).
This systematic error causes the discrepancy between the evolution of line width
with epoch observed and our constant line width in the modelling, and it is
taken into account in our treatment of temporal line-width evolution. Since this
temporal evolution is weak, oxygen and $^{56}$Ni cannot be strongly separated in
the observed SNe and our modelling approach appears to resemble the physical
situation quite well (e.g., a very central $^{56}$Ni distribution would cause a
rapid decrease of line width with time). Thus, we are convinced that our
description of the $^{56}$Ni core of the SNe is sufficiently accurate.

To enable a direct comparison to a parameter which can be obtained without any
detailed modelling, we calculated the half width at half-maximum intensity of the
oxygen doublet [O {\sc i}] $\lambda\lambda$6300, 6364. These two lines are
separated by $\sim$64~\AA, which translates to a velocity of $\sim$3000 km s$^{-1}$. 
With decreasing density the intensity ratio of the two lines will shift from 1:1
to 1:3 \citep{1992ApJ...387..309L,1992SvAL...18..239C}, and in the nebular phase 
the ratio should lie somewhere between 1:2 and 1:3. The error in our characteristic
velocity caused by the superposition of these two lines will therefore be small.

\begin{figure} 
\begin{center}
\includegraphics[width=8cm, clip]{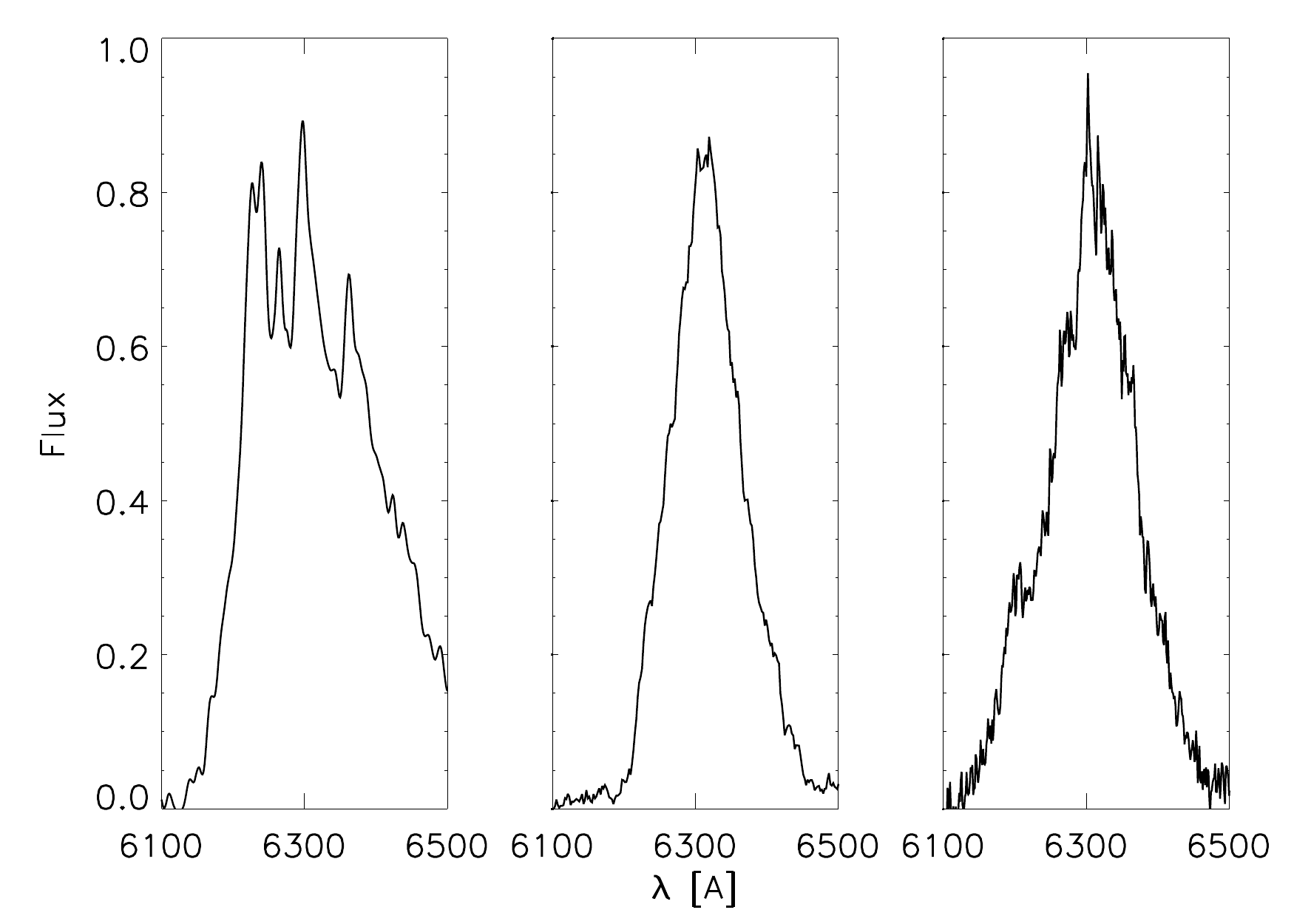}         
\end{center}
\caption{Left: SN 1990B, whose steep blue wing causes a small ratio
of $v_\alpha$ and $v_{50}$. Middle: SN 1983N, which shows an average
ratio of $v_{\alpha}$ and $v_{50}$. Right: SN 2004gq, whose convex
shape causes a large ratio of $v_{\alpha}$ to $v_{50}$, which is
additionally increased by the small notch in the blue wing.}
\label{fig12}
\end{figure}

In Section 2 we already mentioned the advantages of both $v_\alpha$ and
$v_{50}$. While $v_{\alpha}$ is time consuming to compute, $v_{50}$ is less
exact in characterising the velocity of the central ejecta. In Figure \ref{fig5}
one can see a rather linear trend between both characteristic velocities, as
expected. There are, however, some outliers. The SNe in the upper-left corner
(where $v_{\alpha} - 1500$ km s$^{-1} >> v_{50}$) are SNe 1998bw, 2004gq, and
2007I. Both BL-SNe (SN 1998bw, 2007I) have very convex line profiles with a
broad base and a sharp peak. SN 2004gq, the most extreme outlier, also has a
convex shape; in addition, it shows a kink in the line profile, which explains
the large difference between $v_{\alpha}$ and $v_{50}$ for this case. The SNe in
the lower-right region (where $v_{\alpha} - 1500$ km s$^{-1} << v_{50}$) are SNe
1990aa, 1990B, 2005bf, and 2006T. SN 1990aa has a prominent ``spike'' exactly at
half height in the blue wing of the line profile, artificially increasing
$v_{50}$ without affecting $v_{\alpha}$. SN 1990B has a very steep blue wing.
SNe 2005bf and 2006T both have double-peaked [O {\sc i}] profiles with a very
steeply falling blue wing. The width at half height is almost the same as at the
base of the line, which explains the low $v_{\alpha}/v_{50}$ ratio (see Figure
\ref{fig12} for the different line profiles). These examples demonstrate how
$v_{50}$ can give a misleading picture of the characteristic core velocity in
some cases. Hence,  while in general $v_{50}$ seems to be a good proxy of
core velocity, it should not be used if the [O {\sc i}] profile shows broad 
double peaks, spikes, and kinks, or unusually steep or convex wings. 

Both estimates of characteristic velocity are affected by the presence of an
underlying continuum, which can either be the host galaxy or residual continuum
emission form the SN. It is often difficult to distinguish the continuum from
the SN spectrum. We tried to overcome this problem setting a characteristic
minima around the oxygen line to zero flux by removing some linear function from
the spectrum. However, the continuum in most cases is probably not represented
by a linear function. Thus, some additional flux is almost always present in the
region of an emission line, causing an error in the modelling procedure and in
the determination of the half height. Consequently, it is not possible to obtain
a single ``best'' result for a given spectrum. Depending on the quality and
the specific shape of a spectrum, there might be a variety of possible
background subtractions. To cope with this problem, we tried to model the extrema
of what seemed plausible subtractions --- of course a rather arbitrary approach.
We modelled several different SNe in this manner to get a quantitative
estimate of the typical uncertainty and found that an error of $\pm 7$\% should
cover the plausible range (e.g., see Figures \ref{fig2} and \ref{fig3}, where the
differences in characteristic velocity for the models shown are about 5\%).

Another general uncertainty which we cannot quantify is introduced by possible
global asphericities of the SN ejecta. In such a case the projected
velocities might be considerably lower than the actual ejecta velocities, so the 
SN kinetic energy may be underestimated. As long as the ejecta geometry and
inclination are unknown, this problem could not be removed by three-dimensional 
(3D) modelling either.

For 19 SNe ($\sim$35\% of our sample), the shell modelling approach is not
adequate to fit the central parts of the oxygen doublet. This suggests that
at least 35\% of the CC-SNe of our sample might be aspherical in the very centre
\citep[for other explanations, see e.g.][]{2009arXiv0904.4256M}.
\citet{2009arXiv0904.4632T} came to a similar conclusion. Of course,
asphericities cannot be ruled out for the rest of our sample, even if the shell
modelling approach was sufficient to obtain a good fit to the full line profile.
As $v_\alpha$ is dominated by the outer parts of the line profile, a
discrepancy between the model and the observation in the central parts of the
line does not cause large errors in $v_\alpha$.

\section{Results}

Normal SNe of different types seem to have quite similar average core velocities
$v_{\alpha}$: 4402 km s$^{-1}$ ($\sigma_{v} = 403$ km s$^{-1}$) for SNe~IIb (5 
objects), 4844 km s$^{-1}$ ($\sigma_{v} = 935$ km s$^{-1}$) for SNe~Ib (13), and 
5126 km s$^{-1}$ ($\sigma_{v} = 816$ km s$^{-1}$) for SNe~Ic (27).  SNe~Ic have an
average $v_\alpha$ only slightly higher (5\%) than that of SNe~Ib and $\sim$15\%
higher than that of SNe~IIb. SNe BL-Ic have on average a higher $v_\alpha$: 5685 km
s$^{-1}$, and a similar scatter ($\sigma_{v}$ = 824 km s$^{-1}$) to
type Ib and Ic SNe. The sample of SNe~BL-Ic comprises 12 SNe --- 10 of Type Ic, 1 of Type Ib, and 1 
of Type IIb.  Our uncertainties are rather large, but as long as there is no 
systematic over- or underestimate for one of these groups the ratio of their 
averages should be a reliable quantity.

In Section 4 we have shown that if there is a physical correlation between $^{56}$Ni
mass and characteristic core velocity it is weak. There also seems to be
no correlation between total ejecta mass or kinetic energy and core velocity.
There is only a weak correlation between the ratio of total kinetic energy to
total ejecta mass and core velocity. On average, SNe with higher outer ejecta
velocities have higher core velocities, but this trend is weak and shows a
large scatter (see Figure \ref{fig666}). 

\citet{2001AJ....121.1648M} found that SNe~Ic have significantly higher kinetic
energy to mass ratios than SNe~Ib, as measured from the line width at half
maximum. First we compare the SNe contained in \citet{2001AJ....121.1648M} and
our sample. We find that the velocity estimates of \citet{2001AJ....121.1648M}
and $v_{50}$ in this paper agree rather well (to within $\sim$10\%) for most
SNe, consistent with our estimate of the general uncertainty. The difference 
is probably caused by the different sizes of the two samples (2 SNe~Ib and 12
SNe~Ic in \citet{2001AJ....121.1648M}; 13 SNe~Ib and 27 SNe~Ic in this paper). The
listing of line widths at different epochs in \citet{2001AJ....121.1648M} seems
to confirm that line width is rather constant after the nebular phase has been
reached (as discussed above).

\citet{2009arXiv0904.4632T} give the FWHM for 39 CC-SNe, obtained by fitting the 
[O {\sc i}] $\lambda\lambda$6300, 6364 lines with Gaussians. The absolute
values and temporal evolution of these velocities are consistent with our
$v_{50}$ estimates: almost all velocities obtained by Gaussian fitting agree
with $v_{50}$ to 10\% or better, which is within the estimated errors for
$v_{50}$. Velocities obtained with Gaussian fitting are often slightly lower
than $v_{50}$, probably because of the contribution of [O {\sc i}] $\lambda$6364 
to $v_{50}$. The average velocity for SNe~BL-Ic (6 objects) from Gaussian fitting 
is $\sim 3670 \pm 860$ km s$^{-1}$, while the average $v_{50}$ is $\sim 4010
\pm 600$ km s$^{-1}$ for these six objects. Since we expect $v_{50}$ to be
slightly higher than velocities obtained from Gaussian fitting, these estimates
are consistent.

The velocities of SNe~BL-Ic are on average only 10\% higher than those of
regular SNe~Ic. Interestingly, four BL-SNe have core velocities typical of
regular SNe. For two of them (SNe 1997ef, 2003dh) our spectra are extremely
noisy, so the errors could in principle be larger than the estimated
14\% (although we do not expect that), which would allow higher velocities. On
the other hand, for the two other SNe BL-Ic with low core velocities (SNe
1997dq, 2003bg) the spectra are rather good, and high core velocities can 
clearly be ruled out. Since SNe 1997ef and 1997dq are very similar
\citep{2004ApJ...614..858M} in light curve and spectral behaviour, it seems
likely that our estimate for SN 1997ef (which agrees very well with that for SN
1997dq) is also correct. SN 2003dh has a very small $v_\alpha$.

For the GRB-SN 2003dh we might observe a projection effect. Given that a GRB
was detected, we probably observed the SN close to the poles. If the central 
region contained an oxygen-rich disc expanding preferentially near the equatorial
plane, the projected velocity of the [O {\sc i}] line would be much smaller for a
pole-on view. For example, if the disc had an opening angle of $45^\circ$, the
projected velocity would be $\sim 70$\% of the radial ejecta velocity, which
would mean that SN 2003dh would have been observed to have a ``regular''
(relatively high) BL-SN core velocity had it been observed close to the
equator. It is not clear whether the other three BL-SNe with low $v_{\alpha}$
can be explained in a similar way. One would expect roughly one third of all
BL-SNe to be observed close to the poles and two thirds close to the equator,
which is roughly consistent with 4 slow, 7 intermediate, and 1 fast BL-SN (Figure
9 shows that there is a rather steep transition from slow to fast between $40^\circ$ 
and $80^\circ$). 

As the asphericity can be much weaker in the outer layers of the SN than in the
centre, this geometrical interpretation is not in conflict with the early-time
classification. For SN 1998bw, \citet{2007ApJ...668L..19T} have shown that the
classification as BL-SN (from the early spectra) is not affected much by
possible ejecta asphericities.

\begin{figure} 
\begin{center}
\includegraphics[width=8cm, clip]{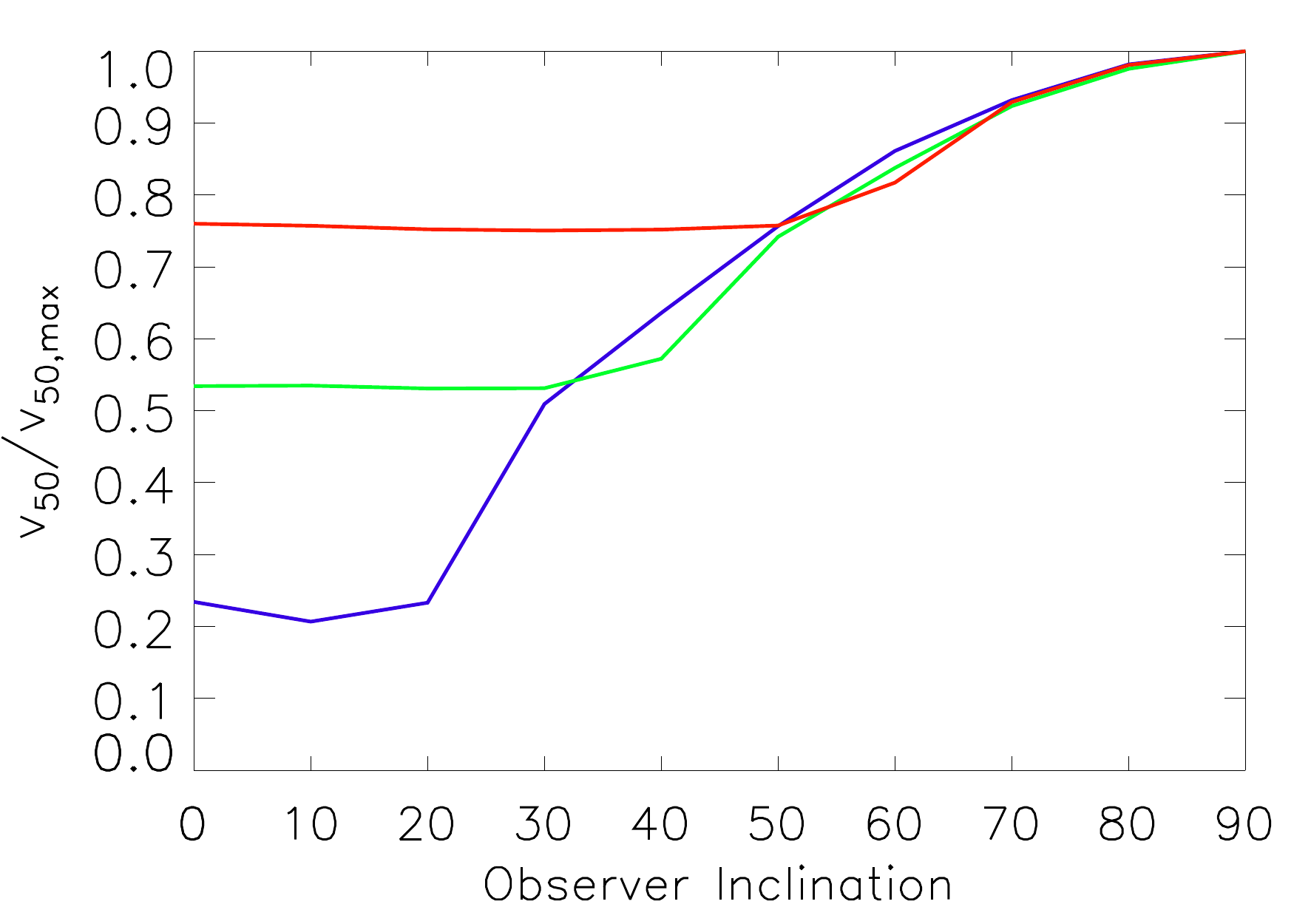}         
\end{center}
\caption{Simulated evolution of normalised $v_{50}$ with observer
  inclination (in $^\circ$, 0$^\circ$ polar, 90$^\circ$ equatorial view) for a $^{56}$Ni/O disc with opening angles of 
  $20^\circ$ (blue, lower line), $40^\circ$ (green, central
  line), and $60^\circ$ (red, upper line). Opening angles of only $20^\circ$
  would cause very low projected velocities, which seem not to be
  observed. Opening angles between $40^\circ$ and $60^\circ$ seem to be able
  to explain the observed range of projected velocities, and are
  consistent with the distribution of high and low
  characteristic BL-SNe (see Figure 3).}
\label{fig11}
\end{figure}

To test whether this scenario is plausible, we computed 3D nebular spectra of
equatorial oxygen discs with different opening angles for several observer
inclinations. We modified the shell nebular code to operate in 3D. For our
simulations we use a spherical grid with $\sim 6000$ cells. We simulate discs of
$^{56}$Ni and oxygen reaching out to 10,000 km s$^{-1}$ with opening angles of
20, 40, and 60 degrees. Varying the observer angle in $10^\circ$ steps from the
pole to the equator, we calculate $v_{50}$ for all three disc opening angles. As
can be seen in Figure \ref{fig11}, oxygen discs with opening angels between $40^\circ$
and $60^\circ$ would be able to explain the observed difference of BL-SNe core
velocities and would also roughly agree with the ratio of low and high
$v_\alpha$ BL-SNe cores. Discs with larger opening angles would result in
smaller variations of the characteristic velocity, while very thin discs
would cause larger differences. Of course, there might be some different
explanations (e.g., different density profiles of the SN progenitors might cause
different ratios of outer and inner velocities), but the geometric
explanation is the most straightforward, especially as asphericities are
expected for a substantial fraction of CC-SNe
\citep[e.g.,][]{2008Sci...319.1220M}.

It is important to note that GRB-SNe 1998bw and 2006aj both have rather high
$v_\alpha$. In the scenario described above, we argued that GRB-SNe might show
low inner ejecta velocities, as we would observe them close from their poles.
For GRB/XRF 060218 there is ongoing discussing whether it is a low-energy GRB
or an XRF. In the latter case the observed X-ray emission would allow no
conclusions about the observer's inclination. Alternatively, and this refers to SN
1998bw as well, SN-GRBs might have larger opening angles than estimated for 
high-redshift GRBs (the determination of GRB opening angles is highly uncertain
anyway, since it is not clear whether the only known method to do so, by measuring 
jet breaks, is reliable). Therefore, these two objects might challenge the
purely geometrical interpretation presented here. Studying the inner
ejecta of future GRB-SNe might shed some light on this issue.

Finally, we mention another interesting object in our sample, SN 2007bi.
Based on our estimate of the $^{56}$Ni mass we could speculate that it is a
pair-instability SN. The nebular spectrum of this SN is very different
from that of the other CC-SNe of our sample. The low core velocity is consistent 
with theoretical predictions that PI SNe produce massive ejecta at moderate
velocities \citep{2005ApJ...633.1031S}.

\section{Summary and Conclusions}
\label{conc}

We estimated $^{56}$Ni masses for 29 SNe for which $^{56}$Ni masses were not
previously known. Their average agrees with the average of 18 CC-SN $^{56}$Ni
masses estimated before rather well, however there might be a small
systematic over-estimate of the $^{56}$Ni mass in this work. Individual
estimates might be quite erroneous given the poor quality of the data set. We then
measured characteristic velocities for 56 CC-SN cores, which range from 3000 km
s$^{-1}$ to 7000 km s$^{-1}$ ($v_{\alpha}$). Several BL-SNe with high-velocity
outer ejecta have high-velocity cores as well, but some BL-SNe do not. We
have shown that this might be due to ejecta asphericities, which are expected
theoretically and might have been detected by different methods before. We found
that the average core velocities of SNe~IIb are slightly lower than core
velocities of SNe~Ib and SNe~Ic. SNe~Ib and Ic have very similar average core
velocities. SNe~IIb show much less variance ($\sigma_v \approx 400$ km s$^{-1}$) of
their core velocities than SNe~Ib and Ic ($\sigma_v \approx 850$ km s$^{-1}$).

There seems to be no strong dependence of core velocity on $^{56}$Ni mass. We
also checked for correlations between total ejecta mass or kinetic
energy and the core velocity and found none. There is only a weak
correlation between the ratio of total kinetic energy to total mass
and the core velocity (BL-SNe have the highest core velocities on average,
but the scatter is so large that there is almost no predictive
power). Therefore, although the total mass
of a SN might be estimated (if the spectrum is flux calibrated), it
is not possible to estimate the SN total kinetic energy from a nebular
spectrum with good accuracy, since the core velocity
seems to correlate with the outer ejecta velocities only weakly. We
will further study the relation between outer and inner ejecta velocities 
in future work.

The uncertainties in our estimates are rather large. They are caused by the general
properties of our method (background subtraction, 
reddening, time evolution of line width,
uncertainty on epoch, $^{56}$Ni distribution, and $^{56}$Ni mass) and are
difficult to improve upon when only limited high-quality data are
available that cover different CC-SNe at several different
epochs. Therefore more and better data is needed.

\section*{Acknowledgements}
This paper presents new observations made with the following
facilities: the European Southern Observatory telescopes (Chile)
obtained from the ESO/ST-ECF Science Archive Facility (Prog ID.
081.D-0173(A), 082.D-0292(A)) and the Gemini Observatory,
which is operated by the Association of Universities for Research in
Astronomy, Inc., under a cooperative agreement with the National
Science Foundation on behalf of the Gemini partnership: the NSF
(United States), the Science and Technology Facilities Council (United
Kingdom), the National Research Council (Canada), CONICYT (Chile), the
Australian Research Council (Australia), Ministério da Ciência e
Tecnologia (Brazil), and Ministerio de Ciencia, Tecnología e Innovación
Productiva (Argentina). A.V.F.'s supernova research has been funded by NSF
grants AST-0607485 and AST-0908886, as well as by the TABASGO Foundation.

\bibliography{pap3}

\begin{thebibliography}{96}
\expandafter\ifx\csname natexlab\endcsname\relax\def\natexlab#1{#1}\fi

\bibitem[{Axelrod}(1980)]{1980PhDT.........1A}
{Axelrod} T.~S., 1980, Ph.D. thesis, AA(California Univ., Santa Cruz.)

\bibitem[{Barkat} et~al.(1967){Barkat}, {Rakavy} \&
  {Sack}]{1967PhRvL..18..379B}
{Barkat} Z., {Rakavy} G., {Sack} N., 1967, Physical Review Letters, 18, 379

\bibitem[{Bartunov} et~al.(1994){Bartunov}, {Blinnikov}, {Pavlyuk} \&
  {Tsvetkov}]{1994A&A...281L..53B}
{Bartunov} O.~S., {Blinnikov} S.~I., {Pavlyuk} N.~N., {Tsvetkov} D.~Y., 1994,
  \aap, 281, L53

\bibitem[{Begelman} \& {Sarazin}(1986)]{1986ApJ...302L..59B}
{Begelman} M.~C., {Sarazin} C.~L., 1986, \apjl, 302, L59

\bibitem[{Blondin} et~al.(2003){Blondin}, {Mezzacappa} \&
  {DeMarino}]{2003ApJ...584..971B}
{Blondin} J.~M., {Mezzacappa} A., {DeMarino} C., 2003, \apj, 584, 971

\bibitem[{Blondin} \& {Calkins}(2008)]{2008CBET.1191....2B}
{Blondin} S., {Calkins} M., 2008, Central Bureau Electronic Telegrams, 1191, 2

\bibitem[{Brown} et~al.(2008){Brown}, {Immler} \& {The Swift Satellite
  Team}]{2008ATel.1403....1B}
{Brown} P.~J., {Immler} S., {The Swift Satellite Team}, 2008, The Astronomer's
  Telegram, 1403, 1

\bibitem[{Burrows} et~al.(2007){Burrows}, {Livne}, {Dessart}, {Ott} \&
  {Murphy}]{2007ApJ...655..416B}
{Burrows} A., {Livne} E., {Dessart} L., {Ott} C.~D., {Murphy} J., 2007, \apj,
  655, 416

\bibitem[{Chugai}(1992)]{1992SvAL...18..239C}
{Chugai} N.~N., 1992, Soviet Astronomy Letters, 18, 239

\bibitem[{Clocchiatti} et~al.(2001){Clocchiatti}, {Suntzeff}, {Phillips}
  et~al.]{2001ApJ...553..886C}
{Clocchiatti} A., {Suntzeff} N.~B., {Phillips} M.~M., et~al., 2001, \apj, 553,
  886

\bibitem[{Clocchiatti} \& {Wheeler}(1997)]{1997thsu.conf..863C}
{Clocchiatti} A., {Wheeler} J.~C., 1997, in { NATO ASIC Proc. 486:
  Thermonuclear Supernovae\/}, edited by P.~{Ruiz-Lapuente}, R.~{Canal},
  J.~{Isern},  863--+

\bibitem[{Clocchiatti} et~al.(1996){Clocchiatti}, {Wheeler}, {Benetti} \&
  {Frueh}]{1996ApJ...459..547C}
{Clocchiatti} A., {Wheeler} J.~C., {Benetti} S., {Frueh} M., 1996, \apj, 459,
  547

\bibitem[{Deng} et~al.(2005){Deng}, {Tominaga}, {Mazzali}, {Maeda} \&
  {Nomoto}]{2005ApJ...624..898D}
{Deng} J., {Tominaga} N., {Mazzali} P.~A., {Maeda} K., {Nomoto} K., 2005, \apj,
  624, 898

\bibitem[{Dimai} \& {Migliardi}(2005)]{2005CBET..300....1D}
{Dimai} A., {Migliardi} M., 2005, Central Bureau Electronic Telegrams, 300, 1

\bibitem[{Dimai} \& {Villi}(2006)]{2006CBET..364....1D}
{Dimai} A., {Villi} M., 2006, Central Bureau Electronic Telegrams, 364, 1

\bibitem[{Drissen} et~al.(1996){Drissen}, {Robert}, {Dutil}
  et~al.]{1996IAUC.6317....2D}
{Drissen} L., {Robert} C., {Dutil} Y., et~al., 1996, \iaucirc, 6317, 2

\bibitem[{Elias} et~al.(1990){Elias}, {Phillips} \&
  {Suntzeff}]{1990IAUC.5080....2E}
{Elias} J., {Phillips} M., {Suntzeff} N., 1990, \iaucirc, 5080, 2

\bibitem[{Elmhamdi} et~al.(2004){Elmhamdi}, {Danziger}, {Cappellaro}
  et~al.]{2004A&A...426..963E}
{Elmhamdi} A., {Danziger} I.~J., {Cappellaro} E., et~al., 2004, \aap, 426, 963

\bibitem[{Filippenko}(1988)]{1988AJ.....96.1941F}
{Filippenko} A.~V., 1988, \aj, 96, 1941

\bibitem[{Filippenko}(1997)]{1997ARA&A..35..309F}
{Filippenko} A.~V., 1997, \araa, 35, 309

\bibitem[{Filippenko} \& {Korth}(1991)]{1991IAUC.5234....1F}
{Filippenko} A.~V., {Korth} S., 1991, \iaucirc, 5234, 1

\bibitem[{Frieman}(2006)]{2006IAUC.8766....1F}
{Frieman} J., 2006, \iaucirc, 8766, 1

\bibitem[{Gabrijelcic} et~al.(1997){Gabrijelcic}, {Benetti} \&
  {Lidman}]{1997IAUC.6535....1G}
{Gabrijelcic} A., {Benetti} S., {Lidman} C., 1997, \iaucirc, 6535, 1

\bibitem[{Galama} et~al.(1999){Galama}, {Vreeswijk}, {van Paradijs}
  et~al.]{1999A&AS..138..465G}
{Galama} T.~J., {Vreeswijk} P.~M., {van Paradijs} J., et~al., 1999, \aaps, 138,
  465

\bibitem[{Gomez} \& {Lopez}(1994)]{1994AJ....108..195G}
{Gomez} G., {Lopez} R., 1994, \aj, 108, 195

\bibitem[{Graham} \& {Li}(2004)]{2004CBET...75....1G}
{Graham} J., {Li} W., 2004, Central Bureau Electronic Telegrams, 75, 1

\bibitem[{Hamuy} et~al.(2009){Hamuy}, {Deng}, {Mazzali}
  et~al.]{2009arXiv0908.1783H}
{Hamuy} M., {Deng} J., {Mazzali} P.~A., et~al., 2009, ArXiv e-prints

\bibitem[{Heger} \& {Woosley}(2005)]{2005IAUS..228..297H}
{Heger} A., {Woosley} S., 2005, in { From Lithium to Uranium: Elemental Tracers
  of Early Cosmic Evolution\/}, edited by V.~{Hill}, P.~{Fran{\c c}ois},
  F.~{Primas}, vol. 228 of { IAU Symposium\/},  297--302

\bibitem[{Hoflich}(1991)]{1991A&A...246..481H}
{Hoflich} P., 1991, \aap, 246, 481

\bibitem[{Itagaki} et~al.(2006){Itagaki}, {Nakano}, {Puckett} \&
  {Toth}]{2006IAUC.8751....2I}
{Itagaki} K., {Nakano} S., {Puckett} T., {Toth} D., 2006, \iaucirc, 8751, 2

\bibitem[{Janka} et~al.(2007){Janka}, {Langanke}, {Marek},
  {Mart{\'{\i}}nez-Pinedo} \& {M{\"u}ller}]{2007PhR...442...38J}
{Janka} H.-T., {Langanke} K., {Marek} A., {Mart{\'{\i}}nez-Pinedo} G.,
  {M{\"u}ller} B., 2007, \physrep, 442, 38

\bibitem[{Jin} et~al.(2007){Jin}, {Cao}, {Bian} et~al.]{2007IAUC.8798....1J}
{Jin} C.~C., {Cao} Y., {Bian} F.-Y., et~al., 2007, \iaucirc, 8798, 1

\bibitem[{Kotake} et~al.(2004){Kotake}, {Sawai}, {Yamada} \&
  {Sato}]{2004ApJ...608..391K}
{Kotake} K., {Sawai} H., {Yamada} S., {Sato} K., 2004, \apj, 608, 391

\bibitem[{Li} \& {McCray}(1992)]{1992ApJ...387..309L}
{Li} H., {McCray} R., 1992, \apj, 387, 309

\bibitem[{Maeda} et~al.(2008){Maeda}, {Kawabata}, {Mazzali}
  et~al.]{2008Sci...319.1220M}
{Maeda} K., {Kawabata} K., {Mazzali} P.~A., et~al., 2008, Science, 319, 1220

\bibitem[{Maeda} et~al.(2007{\natexlab{a}}){Maeda}, {Kawabata}, {Tanaka}
  et~al.]{2007ApJ...658L...5M}
{Maeda} K., {Kawabata} K., {Tanaka} M., et~al., 2007{\natexlab{a}}, \apjl, 658,
  L5

\bibitem[{Maeda} et~al.(2002){Maeda}, {Nakamura}, {Nomoto}, {Mazzali}, {Patat}
  \& {Hachisu}]{2002ApJ...565..405M}
{Maeda} K., {Nakamura} T., {Nomoto} K., {Mazzali} P.~A., {Patat} F., {Hachisu}
  I., 2002, \apj, 565, 405

\bibitem[{Maeda} et~al.(2007{\natexlab{b}}){Maeda}, {Tanaka}, {Nomoto}
  et~al.]{2007ApJ...666.1069M}
{Maeda} K., {Tanaka} M., {Nomoto} K., et~al., 2007{\natexlab{b}}, \apj, 666,
  1069

\bibitem[{Malesani} et~al.(2004){Malesani}, {Tagliaferri}, {Chincarini}
  et~al.]{2004ApJ...609L...5M}
{Malesani} D., {Tagliaferri} G., {Chincarini} G., et~al., 2004, \apjl, 609, L5

\bibitem[{Matheson}(2004)]{2004cetd.conf..351M}
{Matheson} T., 2004, in { Cosmic explosions in three dimensions\/}, edited by
  P.~{H{\"o}flich}, P.~{Kumar}, J.~C. {Wheeler},  351--+

\bibitem[{Matheson} et~al.(2001){Matheson}, {Filippenko}, {Li}, {Leonard} \&
  {Shields}]{2001AJ....121.1648M}
{Matheson} T., {Filippenko} A.~V., {Li} W., {Leonard} D.~C., {Shields} J.~C.,
  2001, \aj, 121, 1648

\bibitem[{Maund} et~al.(2005){Maund}, {Smartt} \&
  {Schweizer}]{2005ApJ...630L..33M}
{Maund} J.~R., {Smartt} S.~J., {Schweizer} F., 2005, \apjl, 630, L33

\bibitem[{Mazzali} et~al.(2002){Mazzali}, {Deng}, {Maeda}
  et~al.]{2002ApJ...572L..61M}
{Mazzali} P.~A., {Deng} J., {Maeda} K., et~al., 2002, \apjl, 572, L61

\bibitem[{Mazzali} et~al.(2004){Mazzali}, {Deng}, {Maeda}, {Nomoto},
  {Filippenko} \& {Matheson}]{2004ApJ...614..858M}
{Mazzali} P.~A., {Deng} J., {Maeda} K., {Nomoto} K., {Filippenko} A.~V.,
  {Matheson} T., 2004, \apj, 614, 858

\bibitem[{Mazzali} et~al.(2006){Mazzali}, {Deng}, {Nomoto}
  et~al.]{2006Natur.442.1018M}
{Mazzali} P.~A., {Deng} J., {Nomoto} K., et~al., 2006, \nat, 442, 1018

\bibitem[{Mazzali} et~al.(2005){Mazzali}, {Kawabata}, {Maeda}
  et~al.]{2005Sci...308.1284M}
{Mazzali} P.~A., {Kawabata} K.~S., {Maeda} K., et~al., 2005, Science, 308, 1284

\bibitem[{Mazzali} et~al.(2007{\natexlab{a}}){Mazzali}, {Kawabata}, {Maeda}
  et~al.]{2007ApJ...670..592M}
{Mazzali} P.~A., {Kawabata} K.~S., {Maeda} K., et~al., 2007{\natexlab{a}},
  \apj, 670, 592

\bibitem[{Mazzali} et~al.(2001){Mazzali}, {Nomoto}, {Patat} \&
  {Maeda}]{2001ApJ...559.1047M}
{Mazzali} P.~A., {Nomoto} K., {Patat} F., {Maeda} K., 2001, \apj, 559, 1047

\bibitem[{Mazzali} et~al.(2007{\natexlab{b}}){Mazzali}, {R{\"o}pke}, {Benetti}
  \& {Hillebrandt}]{2007Sci...315..825M}
{Mazzali} P.~A., {R{\"o}pke} F.~K., {Benetti} S., {Hillebrandt} W.,
  2007{\natexlab{b}}, Science, 315, 825

\bibitem[{Mazzali} et~al.(2008){Mazzali}, {Valenti}, {Della Valle}
  et~al.]{2008Sci...321.1185M}
{Mazzali} P.~A., {Valenti} S., {Della Valle} M., et~al., 2008, Science, 321,
  1185

\bibitem[{McNaught} et~al.(1991){McNaught}, {della Valle} \&
  {Pasquini}]{1991IAUC.5178....1M}
{McNaught} R.~H., {della Valle} M., {Pasquini} L., 1991, \iaucirc, 5178, 1

\bibitem[{Milisavljevic} et~al.(2009){Milisavljevic}, {Fesen}, {Gerardy},
  {Kirshner} \& {Challis}]{2009arXiv0904.4256M}
{Milisavljevic} D., {Fesen} R., {Gerardy} C., {Kirshner} R., {Challis} P.,
  2009, ArXiv e-prints

\bibitem[{Modjaz} et~al.(2008){Modjaz}, {Kirshner}, {Blondin}, {Challis} \&
  {Matheson}]{2008ApJ...687L...9M}
{Modjaz} M., {Kirshner} R.~P., {Blondin} S., {Challis} P., {Matheson} T., 2008,
  \apjl, 687, L9

\bibitem[{Modjaz} et~al.(2009){Modjaz}, {Li}, {Butler}
  et~al.]{2009ApJ...702..226M}
{Modjaz} M., {Li} W., {Butler} N., et~al., 2009, \apj, 702, 226

\bibitem[{Moiseenko} et~al.(2006){Moiseenko}, {Bisnovatyi-Kogan} \&
  {Ardeljan}]{2006MNRAS.370..501M}
{Moiseenko} S.~G., {Bisnovatyi-Kogan} G.~S., {Ardeljan} N.~V., 2006, \mnras,
  370, 501

\bibitem[{Monard}(2006)]{2006IAUC.8666....2M}
{Monard} L.~A.~G., 2006, \iaucirc, 8666, 2

\bibitem[{Monard} et~al.(2004){Monard}, {Quimby}, {Gerardy}
  et~al.]{2004IAUC.8454....1M}
{Monard} L.~A.~G., {Quimby} R., {Gerardy} C., et~al., 2004, \iaucirc, 8454, 1

\bibitem[{Nakamura} et~al.(2000){Nakamura}, {Maeda}, {Iwamoto}
  et~al.]{2000IAUS..195..347N}
{Nakamura} T., {Maeda} K., {Iwamoto} K., et~al., 2000, in { Highly Energetic
  Physical Processes and Mechanisms for Emission from Astrophysical Plasmas\/},
  edited by P.~C.~H. {Martens}, S.~{Tsuruta}, M.~A. {Weber}, vol. 195 of { IAU
  Symposium\/},  347--+

\bibitem[{Nakano} et~al.(1996){Nakano}, {Aoki}, {Kushida}
  et~al.]{1996IAUC.6454....1N}
{Nakano} S., {Aoki} M., {Kushida} R., et~al., 1996, \iaucirc, 6454, 1

\bibitem[{Nakano} et~al.(1997){Nakano}, {Aoki}, {Kushida}
  et~al.]{1997IAUC.6552....1N}
{Nakano} S., {Aoki} M., {Kushida} Y., et~al., 1997, \iaucirc, 6552, 1

\bibitem[{Nomoto} et~al.(1990){Nomoto}, {Shigeyama} \&
  {Filippenko}]{1990BAAS...22.1221N}
{Nomoto} K., {Shigeyama} T., {Filippenko} A.~V., 1990, vol.~22 of { Bulletin of
  the American Astronomical Society\/},  1221--+

\bibitem[{Nomoto} et~al.(1993){Nomoto}, {Suzuki}, {Shigeyama}, {Kumagai},
  {Yamaoka} \& {Saio}]{1993Natur.364..507N}
{Nomoto} K., {Suzuki} T., {Shigeyama} T., {Kumagai} S., {Yamaoka} H., {Saio}
  H., 1993, \nat, 364, 507

\bibitem[{Nugent}(2007)]{2007CBET..929....1N}
{Nugent} P.~E., 2007, Central Bureau Electronic Telegrams, 929, 1

\bibitem[{Parisky} \& {Li}(2007)]{2007CBET.1158....1P}
{Parisky} X., {Li} W., 2007, Central Bureau Electronic Telegrams, 1158, 1

\bibitem[{Pastorello} et~al.(2008){Pastorello}, {Kasliwal}, {Crockett}
  et~al.]{2008MNRAS.389..955P}
{Pastorello} A., {Kasliwal} M.~M., {Crockett} R.~M., et~al., 2008, \mnras, 389,
  955

\bibitem[{Perlmutter} et~al.(1990){Perlmutter}, {Pennypacker}, {Carlson},
  {Marvin}, {Muller} \& {Smith}]{1990IAUC.5087....1P}
{Perlmutter} S., {Pennypacker} C., {Carlson} S., {Marvin} H., {Muller} R.,
  {Smith} C., 1990, \iaucirc, 5087, 1

\bibitem[{Pian} et~al.(2006){Pian}, {Mazzali}, {Masetti}
  et~al.]{2006Natur.442.1011P}
{Pian} E., {Mazzali} P.~A., {Masetti} N., et~al., 2006, \nat, 442, 1011

\bibitem[{Pollas} \& {Maury}(1991)]{1991IAUC.5200....1P}
{Pollas} C., {Maury} A., 1991, \iaucirc, 5200, 1

\bibitem[{Puckett} et~al.(2000){Puckett}, {Langoussis} \&
  {Garradd}]{2000IAUC.7530....1P}
{Puckett} T., {Langoussis} A., {Garradd} G.~J., 2000, \iaucirc, 7530, 1

\bibitem[{Puckett} et~al.(2007){Puckett}, {Orff}, {Madison}
  et~al.]{2007IAUC.8792....2P}
{Puckett} T., {Orff} T., {Madison} D., et~al., 2007, \iaucirc, 8792, 2

\bibitem[{Pugh} et~al.(2004){Pugh}, {Li}, {Manzini} \&
  {Behrend}]{2004IAUC.8452....2P}
{Pugh} H., {Li} W., {Manzini} F., {Behrend} R., 2004, \iaucirc, 8452, 2

\bibitem[{Quimby} et~al.(2004){Quimby}, {Gerardy}, {Hoeflich}
  et~al.]{2004IAUC.8446....1Q}
{Quimby} R., {Gerardy} C., {Hoeflich} P., et~al., 2004, \iaucirc, 8446, 1

\bibitem[{Quimby} et~al.(2007){Quimby}, {Odewahn}, {Terrazas}, {Rau} \&
  {Ofek}]{2007CBET..953....1Q}
{Quimby} R., {Odewahn} S.~C., {Terrazas} E., {Rau} A., {Ofek} E.~O., 2007,
  Central Bureau Electronic Telegrams, 953, 1

\bibitem[{Sahu} et~al.(2009){Sahu}, {Tanaka}, {Anupama}, {Gurugubelli} \&
  {Nomoto}]{2009ApJ...697..676S}
{Sahu} D.~K., {Tanaka} M., {Anupama} G.~C., {Gurugubelli} U.~K., {Nomoto} K.,
  2009, \apj, 697, 676

\bibitem[{Sauer} et~al.(2006){Sauer}, {Mazzali}, {Deng}, {Valenti}, {Nomoto} \&
  {Filippenko}]{2006MNRAS.369.1939S}
{Sauer} D.~N., {Mazzali} P.~A., {Deng} J., {Valenti} S., {Nomoto} K.,
  {Filippenko} A.~V., 2006, \mnras, 369, 1939

\bibitem[{Scannapieco} et~al.(2005){Scannapieco}, {Madau}, {Woosley}, {Heger}
  \& {Ferrara}]{2005ApJ...633.1031S}
{Scannapieco} E., {Madau} P., {Woosley} S., {Heger} A., {Ferrara} A., 2005,
  \apj, 633, 1031

\bibitem[{Schlegel} et~al.(1998){Schlegel}, {Finkbeiner} \&
  {Davis}]{1998ApJ...500..525S}
{Schlegel} D.~J., {Finkbeiner} D.~P., {Davis} M., 1998, \apj, 500, 525

\bibitem[{Schmidt} et~al.(2005){Schmidt}, {Salvo} \&
  {Wood}]{2005IAUC.8472....2S}
{Schmidt} B., {Salvo} M., {Wood} P., 2005, \iaucirc, 8472, 2

\bibitem[{Silverman} et~al.(2009{\natexlab{a}}){Silverman}, {Mazzali},
  {Chornock} et~al.]{2009arXiv0903.4179S}
{Silverman} J.~M., {Mazzali} P., {Chornock} R., et~al., 2009{\natexlab{a}},
  ArXiv e-prints

\bibitem[{Silverman} et~al.(2009{\natexlab{b}}){Silverman}, {Mazzali},
  {Chornock} et~al.]{2009PASP..121..689S}
{Silverman} J.~M., {Mazzali} P., {Chornock} R., et~al., 2009{\natexlab{b}},
  \pasp, 121, 689

\bibitem[{Singer} \& {Li}(2004)]{2004IAUC.8299....1S}
{Singer} D., {Li} W., 2004, \iaucirc, 8299, 1

\bibitem[{Stritzinger} et~al.(2009){Stritzinger}, {Mazzali}, {Phillips}
  et~al.]{2009ApJ...696..713S}
{Stritzinger} M., {Mazzali} P., {Phillips} M.~M., et~al., 2009, \apj, 696, 713

\bibitem[{Stritzinger} et~al.(2006){Stritzinger}, {Mazzali}, {Sollerman} \&
  {Benetti}]{2006A&A...460..793S}
{Stritzinger} M., {Mazzali} P.~A., {Sollerman} J., {Benetti} S., 2006, \aap,
  460, 793

\bibitem[{Takiwaki} et~al.(2009){Takiwaki}, {Kotake} \&
  {Sato}]{2009ApJ...691.1360T}
{Takiwaki} T., {Kotake} K., {Sato} K., 2009, \apj, 691, 1360

\bibitem[{Tanaka} et~al.(2007){Tanaka}, {Maeda}, {Mazzali} \&
  {Nomoto}]{2007ApJ...668L..19T}
{Tanaka} M., {Maeda} K., {Mazzali} P.~A., {Nomoto} K., 2007, \apjl, 668, L19

\bibitem[{Taubenberger} et~al.(2006){Taubenberger}, {Pastorello}, {Mazzali}
  et~al.]{2006MNRAS.371.1459T}
{Taubenberger} S., {Pastorello} A., {Mazzali} P.~A., et~al., 2006, \mnras, 371,
  1459

\bibitem[{Taubenberger} et~al.(2005){Taubenberger}, {Pastorello}, {Mazzali},
  {Witham} \& {Guijarro}]{2005CBET..305....1T}
{Taubenberger} S., {Pastorello} A., {Mazzali} P.~A., {Witham} A., {Guijarro}
  A., 2005, Central Bureau Electronic Telegrams, 305, 1

\bibitem[{Taubenberger} et~al.(2009){Taubenberger}, {Valenti}, {Benetti}
  et~al.]{2009arXiv0904.4632T}
{Taubenberger} S., {Valenti} S., {Benetti} S., et~al., 2009, ArXiv e-prints

\bibitem[{Tokarz} et~al.(1995){Tokarz}, {Garnavich}, {Geller}, {Kurtz},
  {Berlind} \& {Prosser}]{1995IAUC.6271....1T}
{Tokarz} S., {Garnavich} P., {Geller} M., {Kurtz} M., {Berlind} P., {Prosser}
  C., 1995, \iaucirc, 6271, 1

\bibitem[{Tsvetkov}(1986)]{1986PAZh...12..784T}
{Tsvetkov} Y.~D., 1986, Pis ma Astronomicheskii Zhurnal, 12, 784

\bibitem[{Valenti} et~al.(2008{\natexlab{a}}){Valenti}, {Benetti}, {Cappellaro}
  et~al.]{2008MNRAS.383.1485V}
{Valenti} S., {Benetti} S., {Cappellaro} E., et~al., 2008{\natexlab{a}},
  \mnras, 383, 1485

\bibitem[{Valenti} et~al.(2008{\natexlab{b}}){Valenti}, {Elias-Rosa},
  {Taubenberger} et~al.]{2008ApJ...673L.155V}
{Valenti} S., {Elias-Rosa} N., {Taubenberger} S., et~al., 2008{\natexlab{b}},
  \apjl, 673, L155

\bibitem[{Valenti} et~al.(2009){Valenti}, {Pastorello}, {Cappellaro}
  et~al.]{2009Natur.459..674V}
{Valenti} S., {Pastorello} A., {Cappellaro} E., et~al., 2009, \nat, 459, 674

\bibitem[{Williams} et~al.(1996){Williams}, {Martin}, {Germany}, {Schmidt},
  {Stathakis} \& {Johnston}]{1996IAUC.6351....1W}
{Williams} A., {Martin} R., {Germany} L., {Schmidt} B., {Stathakis} R.,
  {Johnston} H., 1996, \iaucirc, 6351, 1

\bibitem[{Wood-Vasey} \& {Chassagne}(2003)]{2003IAUC.8082....1W}
{Wood-Vasey} W.~M., {Chassagne} R., 2003, \iaucirc, 8082, 1

\bibitem[{Woosley} \& {Bloom}(2006)]{2006ARA&A..44..507W}
{Woosley} S.~E., {Bloom} J.~S., 2006, \araa, 44, 507

\end{thebibliography}

\newpage

\section*{Appendix A}

In CC-SNe nebular spectra, the [O {\sc I}] $\lambda\lambda$6300, 6364 doublet often
has a double or even triple peaked emission profile. Recently,
\citet{2009arXiv0904.4256M} studied these peaks for 20 CC-SNe
and found that they are often separated by $\sim$64~\AA. Since this
coincides with the doublet separation, \citet{2009arXiv0904.4256M} concluded that the
multi-peaked line profiles might be caused by some absorption
processes, rather than by ejecta geometry. In our
sample there are 24 SNe with clear multi-peaked [O {\sc I}]
$\lambda\lambda$ 6300, 6364 profiles, and we measured their
peak separation $\Delta$ (see Table 4). 

For SNe 2004ao, 2005kl, 2006T, and 2008ax our results agree with those of
\citet{2009arXiv0904.4256M}. For SN 2003jd, \citet{2009arXiv0904.4256M}
did not measure the separation of the two largest peaks, while we
do, which explains the different value given in this paper. 

Although about 50\% of the
SNe might be
consistent with a separation of 64~\AA (within the uncertainties),
the other 50\% are not. 
For SNe 1985F and 2002ap it seems likely that the second
(very weak) peak is caused by the [O {\sc i}] $\lambda$6364 line, since the
first peak is roughly at 6300~\AA\ while the second is at $\sim$6360~\AA. 
For SNe 1990U, 1991aj, 2000ew, 2003jd, 2004ao, 2004gt, 2006T, 2007I,
2007bi, and 2008ax, the two peaks are of similar strength and are
centered around 6300~\AA. The separation observed in SNe 1990U, 1991aj, 2003jd,
2007I, and 2007bi is clearly not consistent with 64~\AA. 
For the remaining 12 SNe the situation seems even less clear and we refer
to \citet{2009arXiv0904.4256M} for some possible
explanations. 

Interpreting the separation of these peaks as a geometrical effect,
one obtains typical velocities between 2000 km s$^{-1}$ and 4000 km
s$^{-1}$ (1000 km s$^{-1}$ to 2000 km s$^{-1}$ when considering half width) which
is of the same order as the characteristic core velocities measured
in this paper. Therefore
the observed clustering around $\sim$ 3000 km s$^{-1}$ (close to the
64~\AA\ doublet line separation) might be a coincidence caused by the
typical velocity of the SNe cores. 

In conclusion, one can say that ejecta geometry remains an interesting
explanation for split-top line profiles.  

\begin{table}
\begin{tabular}{|l|c|c|}
\hline
SN  & $\Delta$ [\AA] (this work) & $\Delta$ [\AA] \citet{2009arXiv0904.4256M} \\
\hline
1985F & 58 $\pm$ 4&\\
1990B & 65 $\pm$ 10 / 65 $\pm$ 10&\\
1990U & 48 $\pm$ 10&\\
1990aa& 45 $\pm$ 10 / 61 $\pm$ 6&\\
1990aj& 49 $\pm$ 6&\\
1996N & 65 $\pm$ 10&\\
1996aq& 63 $\pm$ 6&\\
2000ew& 59 $\pm$ 6&\\
2002ap& 58 $\pm$ 4&\\
2003bg& 38 $\pm$ 8&\\
2003jd& 100$ \pm$ 20 & 64 $\pm$ 5\\
2004ao& 64 $\pm$ 10 & 65 $\pm$ 3\\
2004dk& 72 $\pm$ 10 &\\
2004gt& 64 $\pm$ 8&\\
2005N & 50 $\pm$ 10&\\
2005bf& 50 $\pm$ 20&\\
2005kl& 65 $\pm$ 10& 65 $\pm$ 4\\
2006ld& 42 $\pm$ 6 / 61 $\pm$ 8&\\
2006T & 70 $\pm$ 20 & 63 $\pm$ 3\\
2007C & 52 $\pm$ 10 / 50 $\pm$ 10&\\
2007I & 45 $\pm$ 10&\\
2007bi& 47 $\pm$ 6&\\
2008D & 45 $\pm$ 10 &\\
2008ax& 61 $\pm$ 4& 64 $\pm$ 1\\
\hline
\end{tabular}
\caption{Peak separation measured in this work and by
  \citet{2009arXiv0904.4256M}. Two values are given if three peaks are
  observed; the separation is then measured from one peak to the
  next. For all SNe but SN 2003jd the estimates
agree rather well. For SN 2003jd the difference is caused by the fact
that \citet{2009arXiv0904.4256M} did not measure the separation of
the two largest peaks. }
\end{table}

\end{document}